\documentclass[prb,twocolumn,showpacs,floatfix]{revtex4-1}
\usepackage{graphics,epsfig,amsfonts,amssymb,amsmath}

\begin{document}
\title{Optical conductivity and Raman scattering of iron superconductors}
\author{B. Valenzuela}
\email{belenv@icmm.csic.es}
\affiliation{Instituto de Ciencia de Materiales de Madrid, ICMM-CSIC, Cantoblanco, E-28049 Madrid (Spain).}
\author{M.J. Calder\'on}
\email{calderon@icmm.csic.es}
\affiliation{Instituto de Ciencia de Materiales de Madrid, ICMM-CSIC, Cantoblanco, E-28049 Madrid (Spain).}
\author{G. Le\'on}
\affiliation{Instituto de Ciencia de Materiales de Madrid, ICMM-CSIC, Cantoblanco, E-28049 Madrid (Spain).}
\author{E. Bascones}
\email{leni@icmm.csic.es}
\affiliation{Instituto de Ciencia de Materiales de Madrid, ICMM-CSIC, Cantoblanco, E-28049 Madrid (Spain).}
\date{\today}
\begin{abstract}
{ 
We discuss how to analyze the optical conductivity and Raman spectra of multi-orbital systems using the velocity and the Raman vertices 
in a similar way Raman vertices were used to disentangle nodal and antinodal regions in cuprates. We apply this method to iron superconductors in the magnetic and non-magnetic states, studied at the mean field level. We find that the anisotropy in the optical conductivity at low frequencies reflects the difference between the magnetic gaps at the $X$ and $Y$ electron pockets. Both gaps are sampled by Raman spectroscopy. We also show that the Drude weight anisotropy in the magnetic state is sensitive to small changes in the lattice structure.}
\end{abstract}
\pacs{75.10.Jm, 75.10.Lp, 75.30.Ds}
\maketitle

\section{Introduction}
Raman and optical conductivity are very useful techniques to analyze the electronic properties of strongly correlated electron systems.\cite{millisrevcond,basovreview,devereauxreview}
Optical conductivity experiments have provided very valuable information on the reorganization of the spectral weight and the opening of gaps in many materials.\cite{basovreview}
In cuprates, the use of different polarizations in Raman scattering has allowed the disentanglement of the different physics of the nodal and the antinodal electronic states.\cite{devereauxreview,nature_sacuto}

The multiband character of iron superconductors complicates the analysis of their Raman\cite{muschlerprb09,devereauxprb10,gallaisprb11,sugai2012} and optical conductivity\cite{Drechsler2008,hu-wang08, boris2009,
pfuner09,qazilbash09,hu-wang09,vanHeumenEPL10,degiorgi10,wu-dressel2010,dong-wang10,lobo2010,
barisic-dressel2010,uchida2011,moon2012,lobo2012} spectra. The five iron 3d orbitals are required for a minimal model to describe these materials. Different interband transitions involve similar energies and contribute to the spectra in the same frequency range. Moreover they start at very small energies\cite{benfatto2011} making it difficult to separate their contribution from the Drude peak in optical conductivity experiments.\cite{qazilbash09,barisic-dressel2010} 

\begin{figure}
\leavevmode
\includegraphics[clip,width=0.4\textwidth]{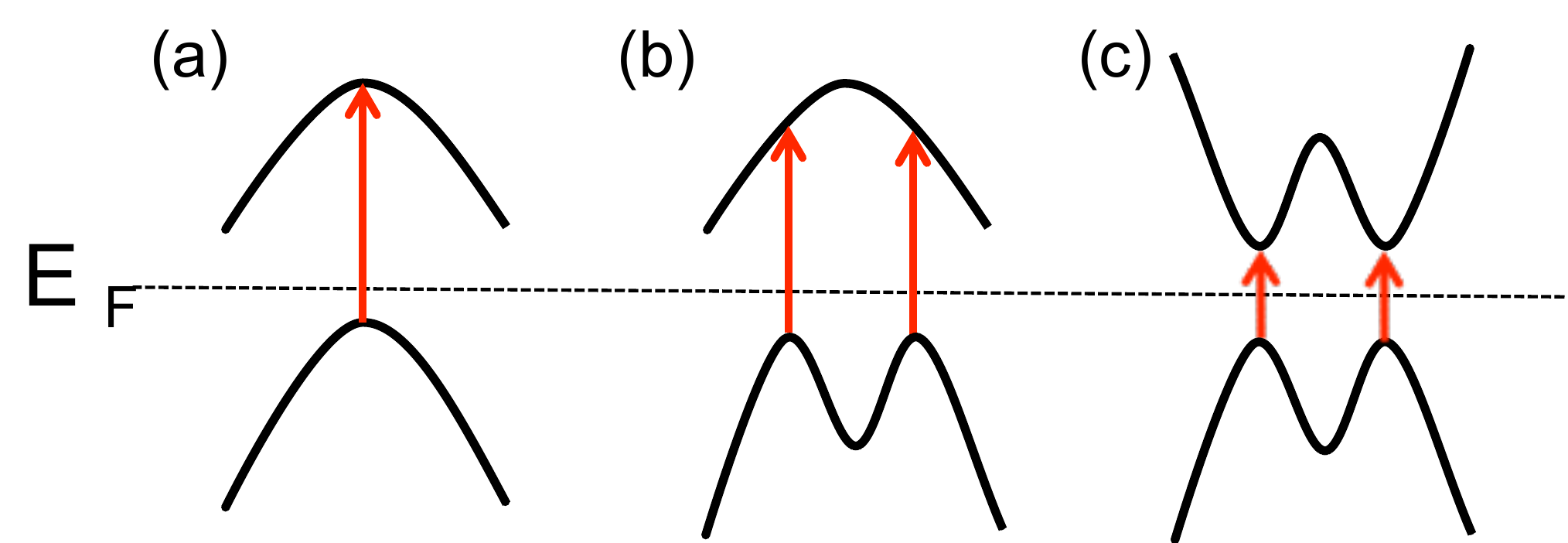}
\vskip -0.2cm
\caption{(Color online) Illustrations of the types of transitions encountered: (a) standard interband transition; (b) transitions involving a non-folded and a folded band; (c) transitions involving two folded anticrossing bands.
}
\label{fig:transitions}
\end{figure}

The difficulties in the interpretation of the spectra are  more pronounced in the magnetic state. When entering the magnetic state the optical conductivity is suppressed at low frequencies and new peaks appear.\cite{hu-wang08,pfuner09,qazilbash09,hu-wang09,dong-wang10,uchida2011,degiorgi10} 
The conductivities along the antiferromagnetic $\sigma_{xx}(\omega)$ and ferromagnetic $\sigma_{yy}(\omega)$ directions show different intensity and peak frequencies.\cite{degiorgi10,uchida2011} The modification of the spectrum and the anisotropy in the magnetic state are visible up to $17000$ cm$^{-1}$, Ref.~[\onlinecite{uchida2011}]. The in-plane resistivity in the magnetic state is also anisotropic. The origin of these anisotropies is not clear yet. \cite{chu10-1,
mazin10,chen2011,uchida2011,uchida2012,degiorgi10,uchida2012-2,
feng2012,chen_deveraux10,nosotrasprl10-2,Fernandes2011,yin11, 
japoneses11,dagottoprb11,lv-phillips2011}
The Raman spectrum in the magnetic state shows signatures and peaks at energies similar to those found in optical conductivity and it has been interpreted in terms of two kinds of electronic transitions: a high energy transition between folded anti-crossed spin density wave bands and a lower energy transition which involves a folded and a non-folded band,\cite{gallaisprb11} see Fig.~\ref{fig:transitions}.  It would be desirable to address the orbital degree of freedom in the interpretation of the optical conductivity and Raman spectroscopies.

\begin{figure*}
\leavevmode
\includegraphics[clip,width=0.9\textwidth]{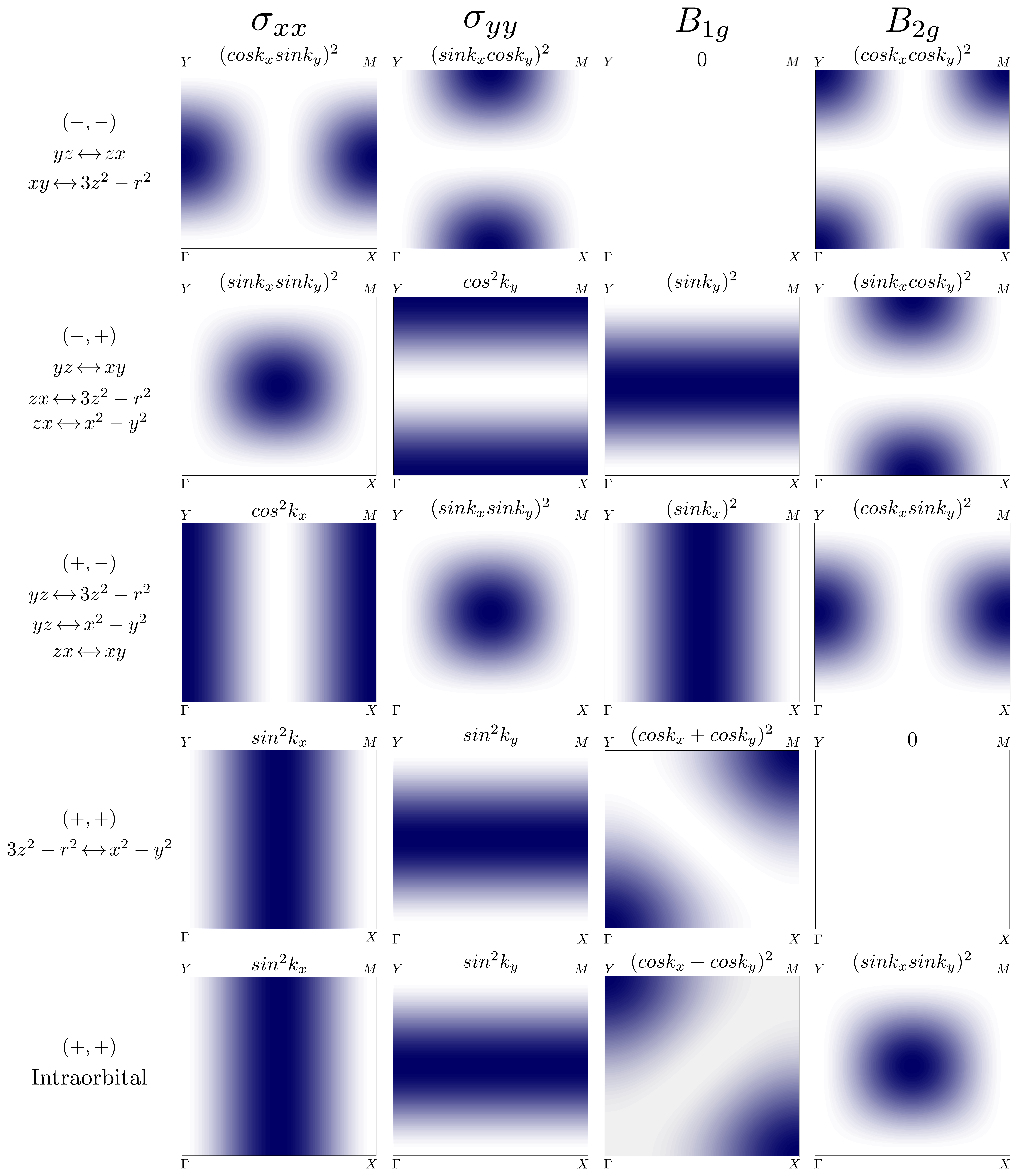}
\vskip -0.2cm
\caption{
(Color online) Leading $\bf k$-dependence of the velocity (optical conductivity)  and Raman squared vertices $\left |v^x_{\mu\nu}\right |^2$, 
$\left |v^y_{\mu\nu}\right|^2$, $\left|\gamma^{B_{1g}}_{\mu\nu}\right|^2$ and 
$\left|\gamma^{B_{2g}}_{\mu \nu}\right|^2$ defined in Eqs.~(\ref{eq:verticevelocidad})-(\ref{eq:b2g}) for iron superconductors for each pair of orbitals involved in an interband transition in the unfolded 1-Fe Brillouin zone corresponding to the tight-binding model in Ref.~[\onlinecite{nosotrasprb09}]. $x$ and $y$ directions point along the Fe-bonds. $B_{1g}$ and $B_{2g}$ are defined in this Brillouin zone and axes. 
The same assignment is used in the non-magnetic and the magnetic state. 
In the case of intraorbital transitions $B_{1g}$ probe the electron 
pockets. Note that a different convention has been used in some Raman 
articles which define the polarizations according to the tetragonal FeAs 
Brillouin zone. Darker regions in each figure correspond to larger values 
of each squared vertex. Combination of optical conductivity and Raman 
scattering spectroscopy allow to probe selectively inter-orbital 
transitions through the whole Brillouin zone. Intraorbital $B_{1g}$ 
squared vertex depends on $\bf k$ as $(A\cos k_x-B\cos k_y)^2$ with the ratio 
$A/B$ dependent on the orbital. $A=B$ valid for $xy$, $x^2-y^2$ and 
$3z^2-r^2$ has been considered in the figure. The $\bf k$-dependence 
changes slightly for $zx$, $yz$, but it is maximal in approximately the 
same regions of the Brillouin zone. The pairs $(\pm,\pm)$ refer to the product of the parities with respect to $x$ and $y$ reflections for each pair of orbitals, see Sec. \ref{sec:vertices} for an illustrative example.
The vertices for the $(+,+)$ $3z^2-r^2 \leftrightarrow x^2-y^2$ transition in $B_{2g}$ symmetry and for the $(-,-)$ transitions in $B_{1g}$ symmetry vanish. Note that the hoppings between $xy$ and $x^2-y^2$ to nearest and next nearest neighbors are zero, so the corresponding vertices cancel within the model considered here. A finite small contribution could arise if hoppings to farther away neighbors were considered. }
\label{fig:optramanvertex}
\end{figure*}

In this paper we discuss how to analyze the optical conductivity and Raman spectrum of multi-orbital systems using the velocity and the Raman vertices.  These vertices depend on the symmetry of the orbitals involved in the interband transitions but do not simply follow the symmetry rules for atomic transitions. They change in $\bf k$-space, see Fig.~\ref{fig:optramanvertex}, reflecting the underlying lattice and can be used to obtain information of the orbitals and the regions of the Brillouin zone (BZ) which contribute to the spectrum in a similar way  Raman vertices were used to disentangle the nodal and antinodal regions in cuprates. We apply this method to iron superconductors in the magnetic and non-magnetic states.  
The information obtained from optical conductivity and Raman spectroscopy are complementary and allow the exploration of all the BZ.

We find that for magnetic moments comparable to the experimental ones the different frequencies at which $\sigma_{xx}$ and $\sigma_{yy}$ peak reflect the magnetic gaps at $Y$ and $X$ electron pockets respectively (in the one-Fe unit cell). For some reconstructed band structures any of these gaps can open below the Fermi level. In this case the interband transition is not allowed affecting the shape of the spectrum. We also show that the Drude weight anisotropy is sensitive to small changes in the lattice structure.

 The article is organized as follows: in Section~\ref{sec:method} we give the expressions for the optical conductivity and the Raman scattering in multi-orbital systems and introduce the respective vertices. Section~\ref{sec:vertices} 
focuses on the Raman and velocity vertices in the case of iron superconductors. In Section~\ref{sec:spectrum} we use the vertices to discuss the optical conductivity and Raman spectra of iron superconductors in the mean-field magnetic and non-magnetic states. The fingerprints in the spectra of the crossover to the orbital differentiation regime which appear in our mean field calculations are analyzed. Section~\ref{sec:Drude} is dedicated to the Drude weight in-plane anisotropy. We end with a discussion of our results and a comparison to experiments in Section~\ref{sec:discussion}.

\newpage

\section{Model and Method}
\label{sec:method}

We consider a multiorbital system with a tight binding Hamiltonian  
\begin{equation}
H_0=\sum_{\bf{i}\bf{j}\sigma\mu\nu}
t_{\bf{i}\bf{j}}^{\mu\nu}
(c_{\bf{i}\mu \sigma}^\dagger c_{\bf {j} \nu\sigma}+h.c)
= \sum_{\bf{k}\sigma\mu\nu}\epsilon_{\mu\nu} (\bf {k})
c^\dagger_{\bf {k} \mu \sigma}c_{\bf {k} \nu \sigma}  \,.
\label{eq:tightbinding}
\end{equation}
$\bf{i}$ and $\bf {j}$ label the lattice sites
connected by the hopping terms, 
$\mu$ and $\nu$ the orbitals, and $\sigma$ the spin. 
$\epsilon_{\mu\nu}(\bf {k})$ is the tight binding  in $\bf k$-space. 

If interactions are local, as considered through all the paper, the coupling between the electrons and the electromagnetic field can be introduced via the Peierls substitution\cite{shastry90,millisrevcond,dagottoreview} in the hopping
$t_{\bf{i}\bf{j}}^{\mu\nu} \to t_{\bf{i}\bf{j}}^{\mu\nu} e^{i\int_{\bf {i}}^{{\bf{i}}+\vec{l}} \vec{A}({\bf {r}}) \cdot d\vec{r}}$.
$\vec{l}=l_x\hat{x}+l_y\hat{y}$ links the neighbors $\bf {i}$ and $\bf {j}$ connected by hopping. Here and in the rest of the paper natural units are used and we take $e=c=\hbar=1$. Assuming that the vector potential $\vec{A} (\bf {r})$ varies more slowly than the lattice length scale one can approximate $\int_{\bf {i}}^{{\bf {i}}+\vec {l}}\vec{A}({\bf {r}}) \cdot d\vec{r} \sim \vec{A} ({\bf{i}} + \vec{l}/2) \cdot \vec{l}$. For small fields we can expand the exponential  to second order  
$e^{i \vec {A} \vec {l}} \sim 1+i \vec{A} \cdot \vec{l}-
\frac{1}{2}(\vec{A}\cdot \vec{l})^2$. Under these approximations and introducing Fourier components, the Hamiltonian in the presence of an electromagnetic field  is

\begin{eqnarray}
\nonumber
 H(\vec{A)} =H(\vec{A}=\vec{0})
&-&\sum_\alpha \sum_{\bf {q}}  j^\alpha_{\bf {q}} A^\alpha_{- \bf{q}} \\ \nonumber
&-&\frac{1}{2}\sum_{\alpha,\beta} \sum_{{\bf {q_1}},{\bf {q_2}}} T^{\alpha \beta}_{\bf {q}_1 +\bf{q}_2} A^\alpha_{- \bf{q}_1} A^\beta_{- \bf{q}_2} ,...
\label{eq:couplingemexp}
\end{eqnarray}
with $\alpha,\beta=x,y$ and 
\begin{eqnarray}
&j^\alpha_{\bf {q}}&=-\sum_{{\bf k}\sigma\mu\nu} \frac{\partial \epsilon_{\mu\nu}(\bf k)}{\partial k_\alpha}c^\dagger_{\bf {k+\frac{q}{2}} \mu\sigma}c_{\bf {k-\frac{q}{2}} \nu \sigma} \, ,
\label{eq:paramag} \\ 
\label{eq:diamag}
&T^{\alpha\beta}_{\bf q}&=-\sum_{{\bf k}\sigma\mu\nu} \frac{\partial ^2\epsilon_{\mu\nu}(\bf {k})}{\partial k_\alpha \partial k_\beta}c^\dagger_{\bf {k+\frac{q}{2}}\mu \sigma}c_{\bf {k-\frac{q}{2}} \nu\sigma}  \, ,\\
&A_{\bf{q}}^\alpha&=\sqrt{\frac{1}{\omega_{\bf{q}}V}}[e^\alpha a_{-\bf{q}}+\bar{e}^\alpha a_{\bf{q}}^\dagger] \, ,
\label{eq:empotential}
\end{eqnarray}
$j^\alpha_{\bf {q}}$ is the paramagnetic current and $T^{\alpha\beta}_{\bf {q}}$ gives rise to the diamagnetic current and enters in the non-resonant Raman response.  $V$ is the volume, $A_{\bf{q}}^\alpha$ the Fourier component of the electromagnetic field, and ${e}^\alpha$ the photon polarization with energy $\omega_{\bf {q}}$ while $\bar{e}^\alpha$ is its complex conjugate.

\subsection{Optical conductivity}
Assuming a single component of the vector potential $A^\alpha_{\bf q}$ and expanding to linear order,
the longitudinal current is given by 
\begin{equation}
J^\alpha_{\bf {q}}  =-\frac{\partial H(\vec{A})}{\partial A^\alpha_{\bf -{q}}} =j^\alpha_{\bf {q}}+\sum_{\bf{q}'} T^{\alpha\alpha}_{\bf {q+q^\prime}}A^\alpha_{-\bf {q}^\prime} \, .
\end{equation}

The contribution of the paramagnetic current to the expected value of the current is given by the Kubo formula. At zero temperature
\begin{equation}
\frac{1}{\omega V}\int_0^\infty e^{i\omega t} \langle \phi_0 \left |[j^{\alpha\dagger}_{\bf {q}}(t),j^\alpha_{\bf {q}}(0)] \right|\phi_0\rangle \, ,
\label{eq:kubo}
\end{equation}
where $\left |\phi_0\rangle \right.$ is the ground state. 
We are interested in the response to a homogeneous  $\bf {q}=0$ 
electric field $A^\alpha_{\bf {q}=0}(\omega)=E^\alpha_{\bf {q}=0}(\omega)/(i\omega -\delta)$ 
with $\delta$ a small parameter, and in particular in 
$\sigma^\prime_{\alpha \alpha}(\omega)$, the real part of the optical 
conductivity, $\sigma_{\alpha \alpha}(\omega)=\sigma^\prime_{\alpha \alpha}(\omega)+i \sigma^{\prime \prime}_{\alpha \alpha}(\omega)$ 
defined as
$J^\alpha_{\bf {q}=0}=\sigma_{\alpha \alpha}(\omega) E^\alpha_{\bf {q}=0}(\omega)$.
After some algebra,\cite{dagottoreview,dagottoprb11}
\begin{equation}
\sigma^\prime_{\alpha\alpha}(\omega)=D_{\alpha}\delta(\omega)
+\frac{\pi}{V}\sum_{m \neq 0} \frac{\left |\langle \phi_0 \left |j^\alpha_{\bf {q}=0} \right |\phi_m\rangle \right|^2}{E_m-E_0} \delta(\omega-(E_m-E_0)),
\label{eq:optcond}
\end{equation}
with the Drude weight given by
\begin{equation}
D_{\alpha}=\pi \langle \phi_0|-T^{\alpha\alpha}_{\bf{q}=0}|\phi_0\rangle-\frac{2\pi}{V}\sum_{m\neq0} \frac{\left|\langle \phi_0 \left |j^\alpha_{\bf {q}=0}\right|\phi_m\rangle \right|^2}{E_m-E_0} \, ,
\label{eq:drude1}
\end{equation}
$E_0$ and $E_m$ are the energies of the ground state $\left | \phi_0 \rangle \right .$ and the excited states $\left |\phi_m\rangle\right.$ respectively. The first and second terms in Eq.~(\ref{eq:drude1}) originate respectively in the diamagnetic and paramagnetic contributions. Eq.~(\ref{eq:optcond}) fulfills the optical sum 
rule\cite{dagottoreview}
\begin{equation}
\int_0^\infty \sigma^\prime_{\alpha \alpha}(\omega) d\omega=\pi \langle \phi_0 \left|-T^{\alpha\alpha}_{\bf {q}=0}\right|\phi_0\rangle \, ,
\end{equation}
that becomes the kinetic energy when hopping is restricted to first nearest neighbors.  

Within the mean-field level used below \cite{nosotrasprl10} the Hamiltonian becomes biquadratic in fermionic operators. Therefore, the eigenstates  $\left|\phi_m \rangle\right.$ can be given in terms of single particle bands and Eqs.~(\ref{eq:optcond}) and (\ref{eq:drude1}) can be written

\begin{eqnarray}
&\sigma^\prime_{\alpha \alpha}(\omega)&=D_{\alpha}\delta(\omega)
+\frac{\pi}{V}\sum_{{\bf k}n \neq n'} \frac{\left |j^\alpha_{n'n}({\bf k}) \right|^2}{\epsilon_{n'}({\bf k})-\epsilon_{n}({\bf k})} 
\nonumber \\&&
\times \theta(\epsilon_{n'}({\bf k})) \theta(-\epsilon_{n}({\bf k})) \delta(\omega-\epsilon_{n'}({\bf k})+\epsilon_{n}({\bf k})),
\label{eq:optcondband}
\end{eqnarray}
\begin{eqnarray}
&D_{\alpha}&=-\pi \sum_{{\bf k}n}t_{nn}^{\alpha\alpha}({\bf k})\theta(-\epsilon_{n}({\bf k})) 
\nonumber \\&&
-\frac{2\pi}{V}\sum_{{\bf k}n \neq n'} \frac{\left |j^\alpha_{n'n}({\bf k})\right|^2}{\epsilon_{n'}({\bf k})-\epsilon_{n}({\bf k})} \theta(\epsilon_{n'}({\bf k})) \theta(-\epsilon_{n}({\bf k}))  \, ,
\label{eq:drude2}
\end{eqnarray}
where $\epsilon_{n}(\bf {k})$ are the band energies, $\theta(\epsilon_{n}({\bf k}))$ the Heaviside step function and 
\begin{eqnarray}
&t_{nn}^{\alpha\alpha}({\bf k})&=\sum_{\mu\nu}\frac{\partial^2 \epsilon_{\mu\nu}({\bf k})}{\partial k_{\alpha}^2}a_{\mu n}^*({\bf k})a_{\nu n}({\bf k}) \, ,
\\
&j_{n'n}^\alpha (\bf {k})&=-\sum_{\mu\nu}\frac{\partial \epsilon_{\mu\nu}({\bf k})}{\partial k_\alpha}a_{\mu n'}^*({\bf k})a_{\nu n}({\bf k}) \, ,
\label{eq:optvertex}
\end{eqnarray}
with $a_{\mu n}({\bf k})$ the rotation matrix between the orbital and the band basis $c_{{\bf{k}}\mu \sigma}^\dagger=\sum_n a_{\mu n}^*({\bf k})d^\dagger_{{\bf {k}}n \sigma}$.

\subsection{Raman response}
The electronic Raman scattering measures the total cross section of the inelastic scattering of electrons.  
It is proportional to the transition rate of scattering an incident $({\bf {q}}_i,\omega_i,\vec{e}_i)$ photon into an outgoing $({\bf {q}}_s,\omega_s,\vec{e}_s)$ state, where ${\bf {q}}_{i,s}$, $\omega_{i,s}$ and $\vec{e}_{i,s}$ label the momentum, frequency and polarization of the incident and scattered photons. The transition rate can be obtained following the Fermi golden rule\cite{devereauxreview,shastry91}
\begin{equation}
\frac{1}{Z}\sum_{I,F}e^{-\beta_B E_I } \left |\langle F \left |M \right|I\rangle\right|^2 \delta(E_F-E_I-\omega) \, ,
\end{equation}
with $Z$ the partition function and $\beta_B=1/K_B T$, being $K_B$ the Boltzmann factor  and $T$ the temperature. $E_I$ and $E_F$ are the initial and final energies of the many electron system, $\omega=\omega_s-\omega_i$ is the transferred energy and $M$ is the effective light scattering between the initial and final state. Neglecting resonant processes\cite{shastry91}
\begin{equation}
\langle F \left |M \right|I \rangle= \sum_{\alpha \beta} e^{\alpha}_i \bar{e}^{\beta}_s \langle F \left |-T^{\alpha \beta}_{\bf q} \right|I\rangle \, ,
\end{equation} 
with ${\bf q}={\bf q}_s-{\bf q}_i$. For the energies involved in the Raman scattering ${\bf q} <<{\bf k}_F$. In the following we take ${\bf q} =0$.

Instead of using $T^{\alpha \beta}_{\bf {q}=0}$ and arbitrary $\alpha$ and $\beta$ polarizations, it is convenient to decompose this matrix element into basis functions of the irreducible point group of the lattice\cite{devereauxreview} according to the polarization of the incident and scattered light $\lambda$= $B_{1g}$, $B_{2g}$, etc and use an effective polarization dependent density matrix 
\begin{equation}
 \rho^\lambda=\sum_{\bf {k}, \sigma \mu\nu}\gamma^\lambda_{\mu\nu} (\bf {k})c^\dagger_{\bf {k} \mu \sigma}c_{\bf {k} \nu \sigma} \, .
\end{equation}
In particular, for $B_{1g}$ and $B_{2g}$ polarizations
\begin{eqnarray}
& \gamma_{\mu\nu}^{B_{1g}}(\bf {k}) &=\frac{\partial^2\epsilon_{\mu\nu}({\bf k})}{\partial k^2_x}-\frac{\partial^2\epsilon_{\mu\nu}({\bf k})}{\partial k^2_y} \, ,
\label{eq:b1g}\\ 
& \gamma_{\mu\nu}^{B_{2g}}(\bf {k}) &= \frac{\partial^2\epsilon_{\mu\nu}({\bf k})}{\partial k_x \partial k_y} \, .
\label{eq:b2g}
\end{eqnarray}
At zero temperature the Raman spectrum for $\lambda$ polarization becomes:
\begin{equation}
S_{\lambda}(\omega) \propto  \sum_m \left|\langle \phi_m \right|{ \rho}^{\lambda}\left|\phi_0\rangle \right|^2 
\delta(E_m-E_0-\omega) \, .
\nonumber
\end{equation}
If, as in the previous subsection, the eigenstates can be expressed in terms of single-particle bands with energies $\epsilon_n(\bf {k})$
\begin{eqnarray}
&S_{\lambda}(\omega)&=\sum_{nn'} \left|\gamma^\lambda_{n'n}({\bf k}) \right|^2 \Theta(-\epsilon_n({\bf{k}}))\Theta(\epsilon_{n'}({\bf {k}})) 
\nonumber \\&& \times \delta(\epsilon_{n'}({\bf k})-\epsilon_n({\bf k})-\omega) \, ,
\nonumber \\
&\gamma^\lambda_{n'n}({\bf k})&=\sum_{\mu\nu}\gamma^\lambda_{\mu\nu}({\bf {k}}) a_{\mu n'}^*({\bf{k}}) a_{\nu n}({\bf{k}}) \, .
\label{eq:eqraman}
\end{eqnarray}
$S_\lambda (\omega)$ is related\cite{kuzmanybook} to $\chi^{''}_{\lambda}(\omega)$, the imaginary part of the effective polarization density correlation function $\chi_\lambda (\omega)$,
\begin{equation}
\chi_\lambda(\omega)=i\int_0^\infty dt e^{-i\omega t}\langle [\rho^\lambda(t),\rho^\lambda(0)]\rangle \, ,
\end{equation}
discussed below by:
\begin{equation}
\chi^{''}_{\lambda}(\omega)=\pi \left [S_{\lambda}(\omega)-S_{\lambda}(-\omega) \right ]\, .
\end{equation}
At zero temperature and zero scattering rate $S_\lambda(\omega) \propto \chi^{''}_\lambda(\omega)$. 
In our calculations we broaden the delta functions with a small scattering rate $\Gamma=20$ meV. With such $\Gamma$, the proportionality relation between $\chi^{''}_\lambda(\omega)$ and  $S_\lambda(\omega=0)$  only fails at $\omega \sim 0$. $\chi^{''}_\lambda(\omega=0)=0$ while $S_\lambda(\omega=0)$ acquires a small finite value.  However, the qualitative features of the spectra are not affected.

\begin{figure*}
\leavevmode
\includegraphics[clip,width=0.47\textwidth]{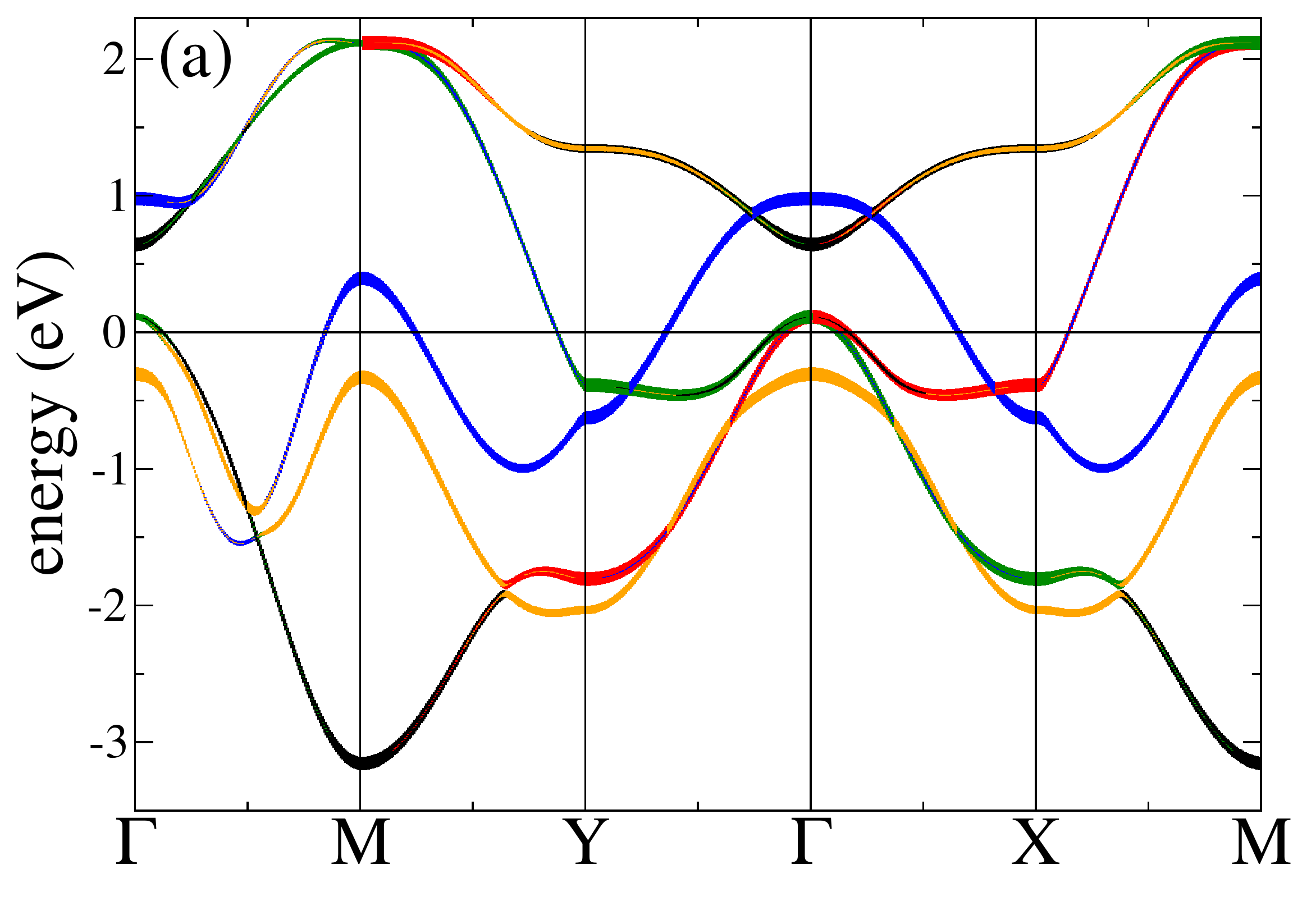}
\includegraphics[clip,width=0.47\textwidth]{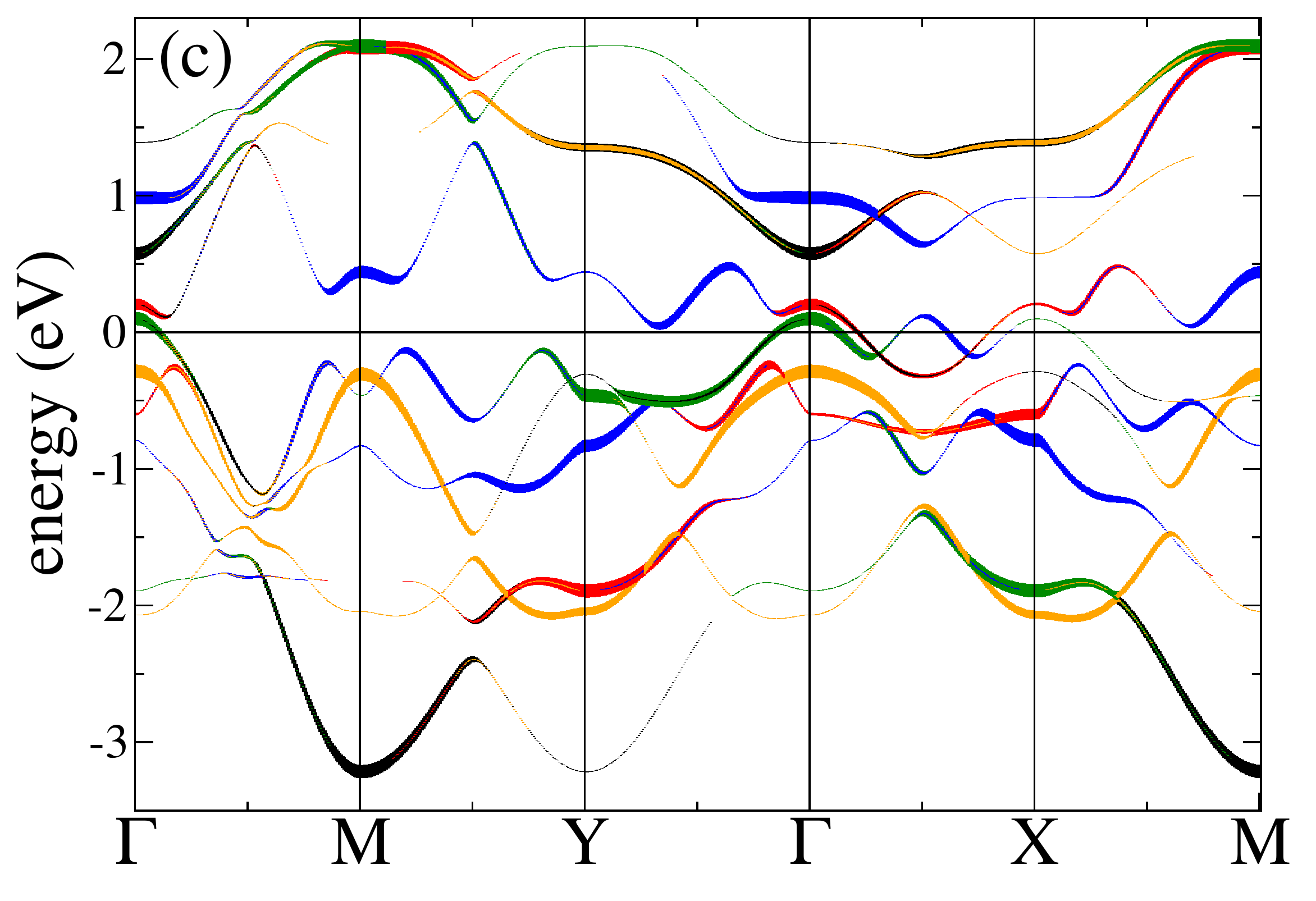}
\includegraphics[clip,width=0.47\textwidth]{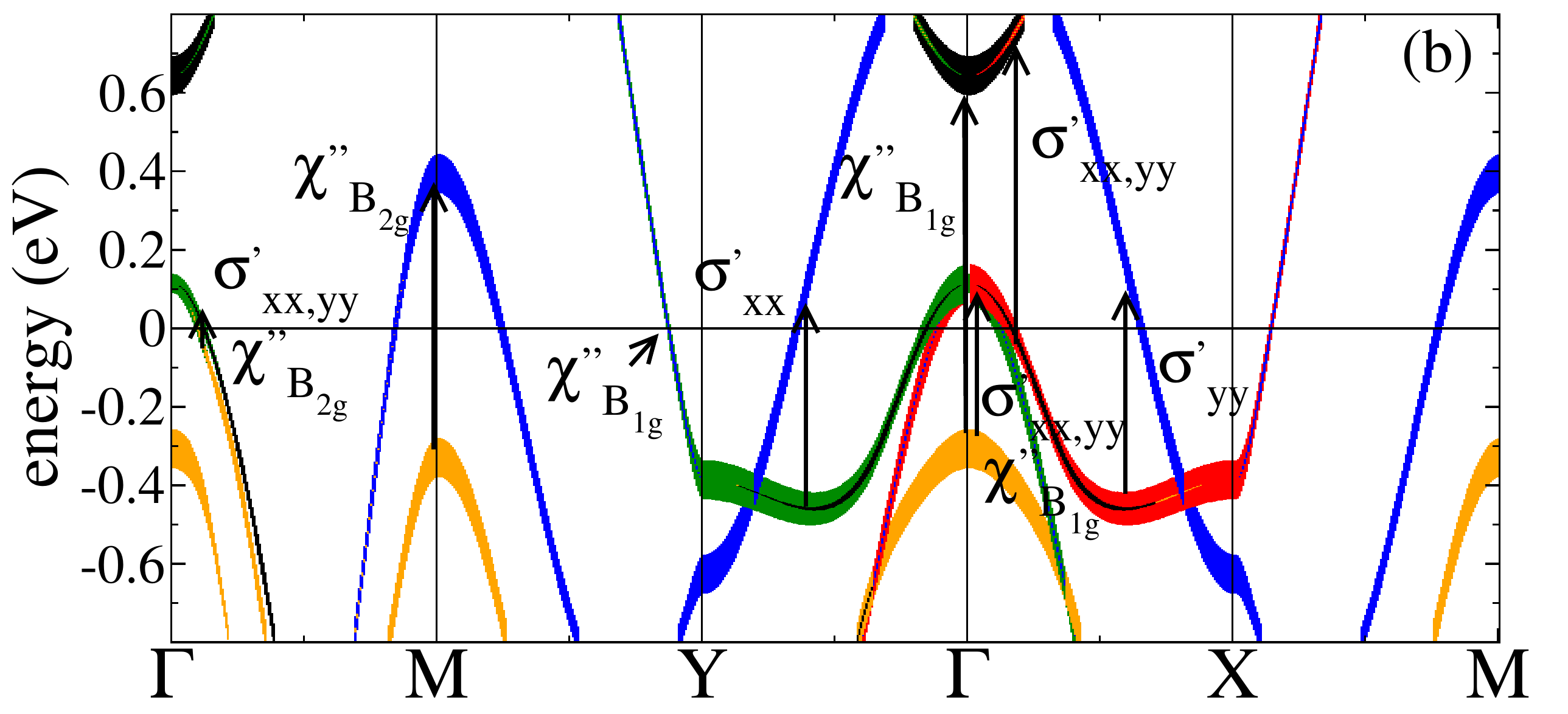}
\includegraphics[clip,width=0.47\textwidth]{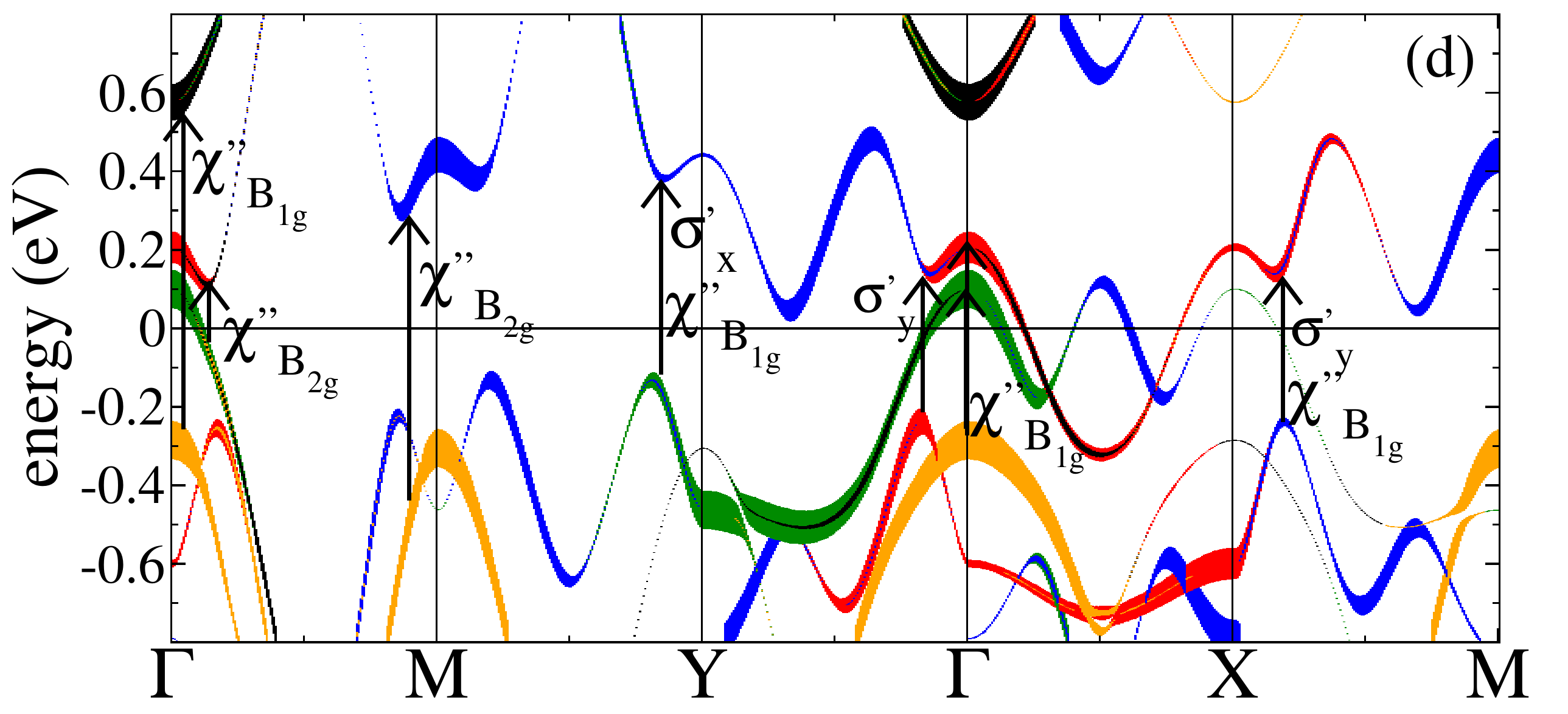}
\vskip -0.2cm
\caption{(Color online) Upper figures: Energy bands in the  non-magnetic (a) and $(\pi,0)$ antiferromagnetic state (c) for $ U=J_H=0$ and $U=1.6$ eV and $J_H=0.25 U$ respectively. Linewidths and colors reflect the orbital content $yz$=red, $zx$=green, $xy$=blue, $3z^2-r^2$=orange and $x^2-y^2$=black. Note that not all the orbital content is visible due to overlapping curves. Lower figures: Zoom of the band structures in the upper figures in the energy region close to the Fermi level showing the optical and Raman transitions at low energy discussed in the text.}
\label{fig:bandsU1p6Jp25}
\end{figure*}

\subsection{Raman and velocity vertices}
According to Eqs.~(\ref{eq:optcondband}), (\ref{eq:optvertex}) and (\ref{eq:eqraman}) the optical conductivity and Raman spectrum are given in terms of the quantities $\left|j_{nn'}^\alpha(\bf {k}) \right|^2$ and $\left|\gamma_{nn'}^\lambda(\bf {k})\right|^ 2$. Once the square is expanded in orbitals the optical conductivity involve terms 
\begin{equation}
\sum_{\mu \mu'\nu \nu'} \frac{\partial \epsilon_{\mu \nu}(\bf {k})}{\partial k_\alpha} \frac{\partial \epsilon^*_{\mu' \nu'}(\bf {k})}{\partial k_\alpha} a_{\mu n}^*({\bf k})a_{\mu' n}({\bf k})a_{\nu n'}({\bf k})a_{\nu' n'}^*({\bf k}). 
\nonumber
\end{equation}
Similar terms appear in the calculation of the Raman spectrum. The transition probability between $n$ and $n'$ has a very complex $k$-structure which depends on the interactions through the $a_{\mu n}$ operators, making very difficult the analysis of the spectra.

In order to get a simple picture of the transitions, one can neglect the crossed terms with $\mu \neq \mu'$ or $\nu \neq \nu'$ and focus on the terms with $\mu=\mu'$ and $\nu=\nu'$. Within this approximation the probabilities for an interband transition between bands $n$ and $n'$ in optical conductivity  $\left|j_{nn'}^\alpha \right|^2$ and Raman spectroscopy 
$\left |\gamma_{nn'}^\lambda \right|^ 2$  in Eqs.~(\ref{eq:optcondband}) and (\ref{eq:eqraman}) are
\begin{eqnarray}
\left|j_{nn'}^\alpha(\bf {k})\right|^2 & \sim & \sum_{\mu\nu}\left|v^\alpha_{\mu\nu}(\bf {k})\right|^2 \left|a_{\mu n}({\bf k})\right|^2 \left|a_{\nu n'}({\bf k})\right|^2 \, ,\label{eq:velocity}\\
\left|\gamma_{nn'}^\lambda(\bf {k})\right|^2 & \sim & \sum_{\mu\nu}\left|\gamma^\lambda_{\mu\nu}(\bf {k})\right|^2 \left|a_{\mu n}({\bf k})\right|^2 \left |a_{\nu n'}({\bf k})\right|^2 \, .
\label{eq:gamma}
\end{eqnarray}
Here $\left|a_{\mu n}({\bf k})\right|^2$ and $\left|a_{\nu n'}({\bf k})\right|^2$ give the spectral weight of orbitals $\mu$ and $\nu$ in the bands $n$ and $n'$ involved in the transition. For each transition between bands $n$ and $n'$, this approximation only keeps the weights corresponding to $\mu$ orbital-density on band $n$ $|a_{\mu n}|^2$ to $\nu$ orbital-density on band $n’$ $|a_{\nu n’}|^2$, summing to all pairs of orbitals. The  velocity $\left|v^\alpha_{\mu \nu}(\bf{k})\right|^2$ and Raman $\left|\gamma^\lambda_{\mu \nu}(\bf{k})\right|^2$ squared vertices are given by 
\begin{eqnarray}
& \left|v_{\mu\nu}^{\alpha}(\bf {k})\right|^2 &=\left|\frac{\partial\epsilon_{\mu\nu}({\bf k})}{\partial k_\alpha}\right|^2 \, ,
\label{eq:verticevelocidad}\\ 
& \left|\gamma_{\mu\nu}^{B_{1g}}(\bf{k})\right|^2 &=\left|\frac{\partial^2\epsilon_{\mu\nu}({\bf k})}{\partial k_x^2}-\frac{\partial^2\epsilon_{\mu\nu}({\bf k})}{\partial k_y^2}\right|^2 \, ,
\label{eq:b1g}\\ 
& \left|\gamma_{\mu\nu}^{B_{2g}}(\bf {k})\right|^2 &= \left|\frac{\partial^2\epsilon_{\mu\nu}({\bf k})}{\partial k_x \partial k_y}\right|^2 \, ,
\label{eq:b2g}
\end{eqnarray}
where $\alpha=x,y$ and we have focused on the $\lambda= B_{1g},B_{2g}$ Raman polarizations. 

Besides the corresponding spectral weights $|a_{\mu n}|^2$ and $|a_{\nu n’}|^2$, the probabilities of a transition from the orbital component $\mu$ in band $n$ to the orbital component $\nu$ in band $n'$ in a Raman or optical conductivity experiment are respectively weighted  by the squared vertices $\left|\gamma^\lambda_{\mu\nu}(\bf {k})\right|^2$ and $\left|v^\alpha_{\mu\nu}(\bf {k})\right|^2$. These vertices depend on $\bf {k}$ through the underlying lattice via  $\epsiloṇ_{\mu\nu}(\bf {k})$. While the vertices $ \left|v_{\mu\nu}^{\alpha}(\bf {k})\right|^2$ and $\left|\gamma_{\mu\nu}^{B_{1g}}(\bf{k})\right|^2$ depend on the symmetry of the orbitals, they cannot be deduced from simple arguments involving atomic optical transitions, valid only at $\Gamma$.

Therefore, knowledge of the tight binding dispersion  $\epsilon_{\mu\nu} (\bf {k})$ in the orbital basis helps to identify which transitions contribute to the optical conductivity and Raman spectrum. The
band structure of the interacting Hamiltonian in the normal or ordered state is also required for the interpretation of the spectrum via the orbital spectral weight $\left|a_{\nu n}\right|^2$ and the band energies $\epsilon_n(\bf {k})$.

We emphasize that the approximation in Eqs. [\ref{eq:velocity}-\ref{eq:b2g}] is used only in the interpretation of the spectra that is calculated using the full expressions in Eqs. \ref{eq:optcondband} and \ref{eq:eqraman}.

\section{Vertices in iron superconductors}
\label{sec:vertices} 
Previous expressions are valid for a multiorbital system with local interactions. In the rest of the paper we focus on the case of iron superconductors. Unlike otherwise indicated we consider the five-orbital tight binding model introduced by the authors in Ref.~[\onlinecite{nosotrasprb09}], where the five orbitals refer to the 3d iron orbitals $yz$, $zx$, $xy$, $3z^2-r^2$, $x^2-y^2$. The model and the orbital directions are defined in the one-Fe unit cell, with $x$ and $y$ along the Fe-Fe bonds. Hopping among the orbitals is restricted to first and second neighbors and includes both indirect hopping mediated by As as well as direct hopping between Fe orbitals. 
Indirect hopping depends on the angle $\alpha_{Fe-As}$ formed by the Fe-As bond and the Fe plane. The tight-binding model used allows the modification of this angle. Except otherwise indicated the results are given for a regular tetrahedron with $\alpha_{Fe-As}=35.3^\circ$ with a non-magnetic band structure as plotted in Fig.~\ref{fig:bandsU1p6Jp25} (a). 
Similar results are expected for other five-orbital models discussed in the literature, whenever defined using the same orbitals and unit cell. 

The five orbitals result in 15 different squared vertices ($5$ intraorbital and $10$ interorbital) for each of the spectroscopies discussed in this article: $\sigma^\prime_{xx}$, $\sigma^\prime_{yy}$, $\chi^{\prime \prime}_{B_{1g}}$ and $\chi^{\prime \prime}_{B_{2g}}$. Keeping only the leading $\bf k$-dependence, one can  group the squared vertices according to the symmetry of the product of the orbital wavefunctions involved in the transition with respect to $x$ and $y$ reflections.  For example, $yz$ orbital is even (odd) with respect to $x$ ($y$) while $xy$ is odd with respect to both $x$ and $y$. Therefore, the product $yz \cdot xy$ is odd (even) with respect to $x$ ($y$), summarized in $(-,+)$. The pairs $yz, xy$ and $zx, 3z^2-r^2$ have the same product parity $(-,+)$. For both terms the derivative involved in the vertex of $\sigma_y$ is $\partial\epsilon_{\mu,\nu}/ \partial k_y=2 i \cos k_y (t^y_{\mu,\nu} - 2 \tilde t_{\mu,\nu}\cos k_x)$ with $t$ and $\tilde t$ the hoppings for the corresponding orbitals to nearest and next nearest neighbors respectively.\cite{nosotrasprb09} The leading $k$-dependence of the squared vertex for $\sigma_y$ is $\cos^2 k_y$ for the orbital pairs with product parity $(-,+)$.

The leading $\bf k$-dependence of the vertices is shown in Fig. \ref{fig:optramanvertex}, where the darkest color emphasizes the region of the BZ with largest square vertex and the product symmetry is given in parenthesis ($\pm,\pm)$ with $+$ and $-$ referring respectively to even and odd.

In Fig.~\ref{fig:optramanvertex} the vertices for the  $(-,-)$ transitions in $B_{1g}$ symmetry vanish.
The $(+,+)$ symmetry group includes both the transition $3z^2-r^2 \leftrightarrow x^2-y^2$ and the intraorbital ones. The optical conductivity velocity vertices of these transitions have the same leading $\bf k$-dependence but there are some differences in the Raman case. While the squared $B_{2g}$ vertex vanishes for the $3z^2-r^2 \leftrightarrow x^2-y^2$ transition it depends as $(\sin k_x \sin k_y)^2$ for intraorbital transitions. The $B_{1g}$ squared vertex goes like $(A\cos k_x-B\cos k_y)^2$ with $A$ and $B$  orbital dependent. $A=-B$ for the  $3z^2-r^2 \leftrightarrow x^2-y^2$ transition while $A=B$ for intraorbital transitions involving $xy$, $x^2-y^2$ or $3z^2-r^2$. This results in a  different $\bf k$-dependence of the squared vertices with maximum and minimum values in different regions in $\bf k$-space. For $zx$ and $yz$ orbitals $A \neq B$ but the ${\bf k}$-dependence is similar to that shown for $A=B$.

The $x^2-y^2$ intraorbital Raman squared vertices depend on $\bf {k}$ as $(\cos k_x-\cos k_y)^2$ for $B_{1g}$ and $(\sin k_x \sin k_y)^2$ for $B_{2g}$. This dependence was widely used in the analysis of the cuprates Raman spectrum and allowed to separate the physics of the antinodal region sampled by $B_{1g}$ from that of the nodal region measured by $B_{2g}$.  In a similar way as done in the cuprates for the $x^2-y^2$ intraorbital transitions, the contributions to the spectrum of a transition between two bands weighted by two given orbitals will be larger when the corresponding vertex is large in the region of $\bf {k}$-space which satisfies the energy conservation and this can be used to analyze the experimental and calculated spectrum. The region of the BZ weighted for transitions between two orbitals is different for the optical conductivity and Raman vertices, being possible to cover practically all the BZ for every transition by studying $\sigma^{\prime}_{xx}$, $\sigma^\prime_{yy}$, $\chi ^{\prime \prime}_{B_{1g}}$ and $\chi^{\prime \prime}_{B_{2g}}$. Note that due to the multiorbital character $B_{1g}$ ($B_{2g}$) can sample different regions of the BZ besides the antinodal (nodal) region.

\begin{figure}
\leavevmode
\includegraphics[clip,width=0.45\textwidth]{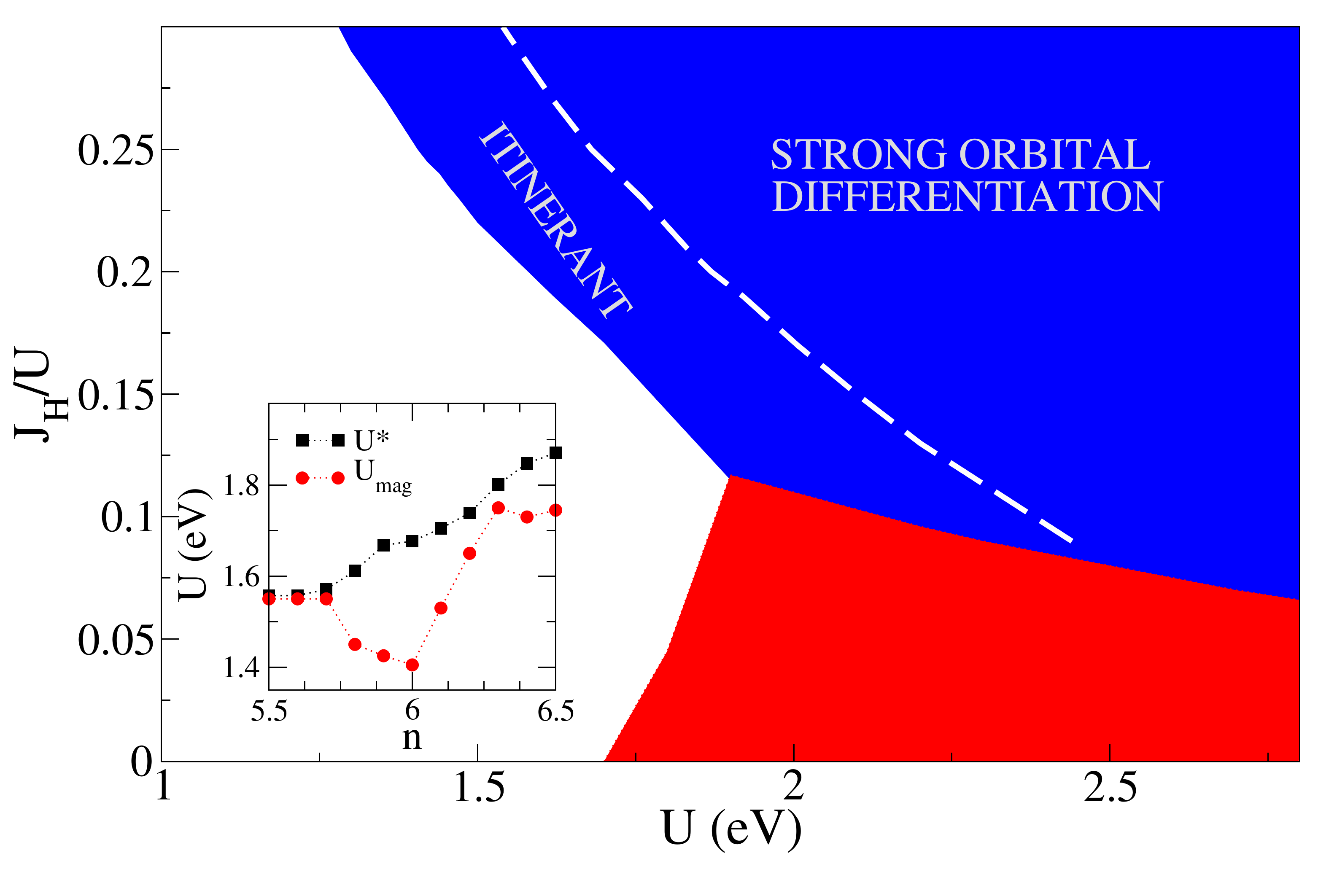}
\vskip -0.2cm
\caption{(Color online) Main figure: $(\pi , 0)$ magnetic phase of the undoped ($6$ electrons in $5$ d orbitals) system as a function of the interactions $U$ and $J_H/U$. The white area is the non-magnetic region. The blue and red areas are magnetic with a high moment (parallel orbital moments) and a low moment (antiparallel orbital moments) state respectively, studied in Ref.~[\onlinecite{nosotrasprl10}]. 
The dashed line separates the itinerant and the strongly orbital differentiated magnetic regions, see text and Ref.~[\onlinecite{nosotrasprb12-2}] for discussion. Inset: Doping dependence of $U_{mag}$ ($U^*$), the interactions at which antiferromagnetism (strong orbital differentiation) appear for $J_H=0.25 U$, reproduced from Ref.~[\onlinecite{nosotrasprb12-2}].}
\label{fig:phasediagram}
\end{figure}

\section{Optical conductivity and Raman spectrum of iron superconductors}
\label{sec:spectrum}
\subsection{Phase diagram and band reconstruction}
To discuss the spectrum of iron superconductors in the magnetic and non-magnetic states, we consider a five orbital interacting Hamiltonian with local interactions. Assuming rotational invariance, the interactions can be written in terms of two parameters: the intraorbital interaction $U$ and the Hund's coupling $J_H$, see Ref.~[\onlinecite{nosotrasprl10}] for details. Except otherwise indicated we assume electron filling $n=6$, as in undoped materials, and use our tight binding model in Ref.~[\onlinecite{nosotrasprb09}] with squared vertices as given in the previous section. The Hamiltonian is treated at the mean field level.

The mean field $(\pi,0)$ magnetic phase diagram as a function of the interactions $U$ and $J_H/U$ has been discussed previously~\cite{nosotrasprl10,nosotrasprb12-2} and is reproduced in Fig.~\ref{fig:phasediagram} for clarity, see also Refs.~[\onlinecite{nosotrasprb12}] and [\onlinecite{dagotto-arpes}]. The white area is the non-magnetic state and the red area corresponds to a low moment state with antiparallel orbital moments violating Hund's rule.\cite{nosotrasprl10} The blue area corresponds to a magnetic state with parallel orbital moments. In this phase there is a crossover from an itinerant to a  strong orbital differentiation regime represented in Fig.~\ref{fig:phasediagram} by a white dashed line. In the strong orbital differentiation regime $xy$ and $yz$ are half-filled gapped states while $zx$, $3z^2 -r^2$ and $x^2 -y^2$ are itinerant with a finite density of states at the Fermi level.~\cite{nosotrasprb12-2}

With hole doping the system becomes more correlated \cite{nosotrasprb12-2,liebsch2010,Werner2012,imadaPRL2012,si2012-2}
as the average orbital filling approaches half-filling.  
As shown in the inset of Fig.~\ref{fig:phasediagram} the interaction $U^*$ at which the system enters into the orbital differentiated regime decreases with hole doping while  $U_{mag}$, the interaction at which antiferromagnetism appears,  is non-monotonous.\cite{nosotrasprb12-2}  

\begin{figure}
\leavevmode
\includegraphics[clip,width=0.23\textwidth]{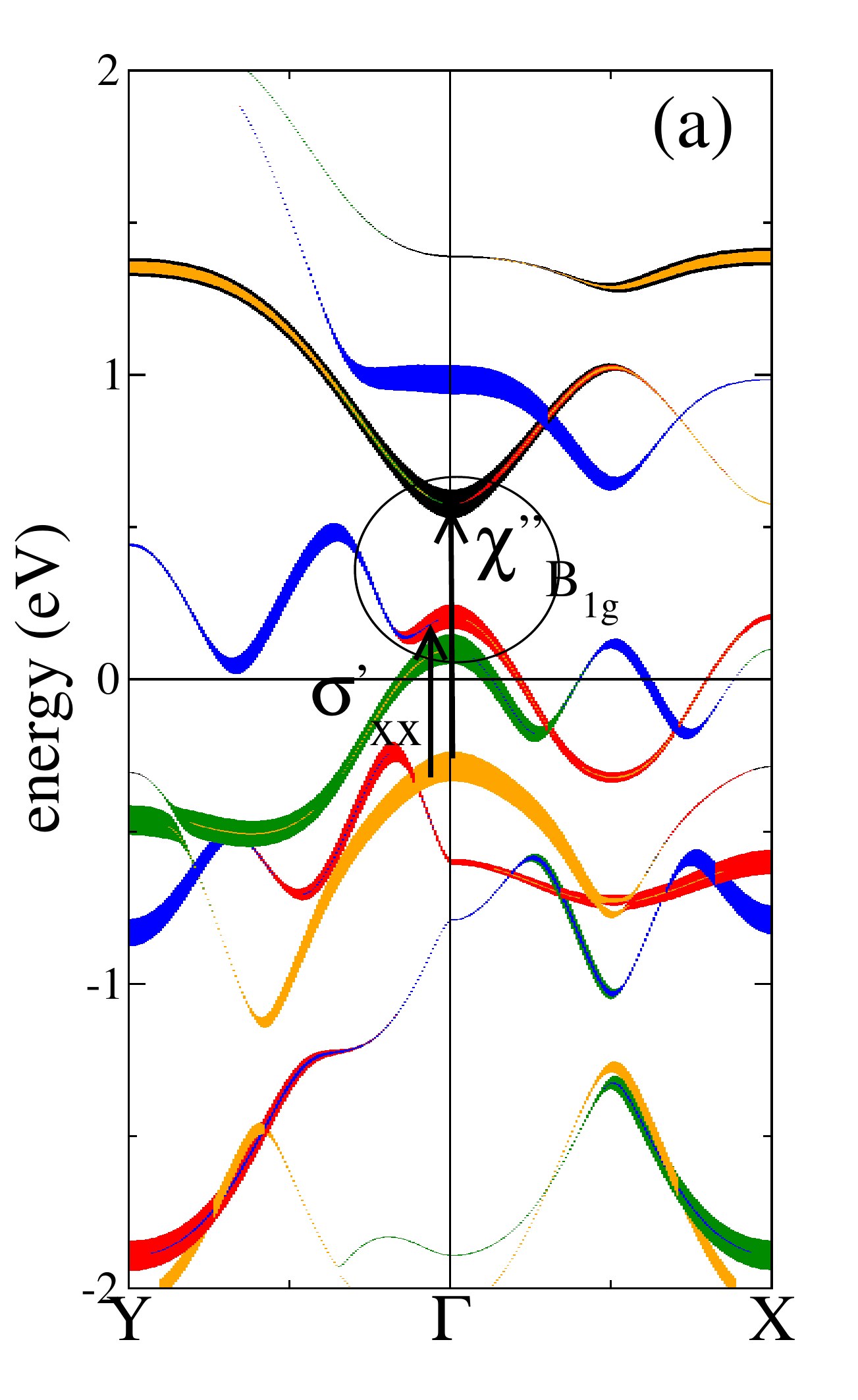}
\includegraphics[clip,width=0.23\textwidth]{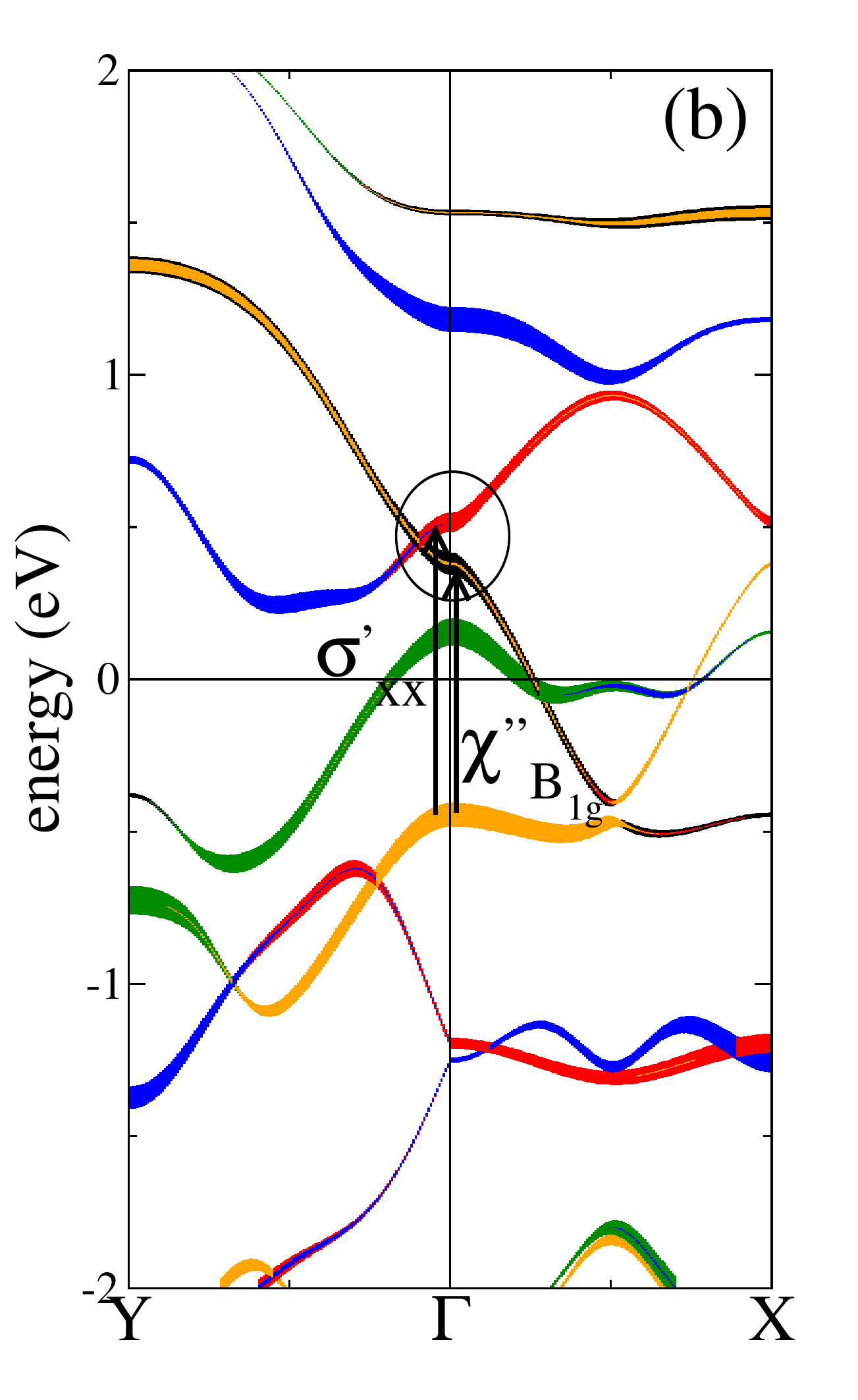}
\vskip -0.2cm
\caption{(Color online) Band structure in the $Y$-$\Gamma$-$X$ directions in the antiferromagnetic state corresponding to $J_H=0.25 U$ with (a) $U=1.6$ eV and (b) $U=1.8$ eV, in the itinerant and orbital differentiated regions respectively. Linewidths and colors reflect the orbital content $yz$=red, $zx$=green, $xy$=blue, $3z^2-r^2$=orange and $x^2-y^2$=black. The circles highlight the shift of the $yz$ orbital to higher energies, driving $yz$ to a half-filled gapped state, and $x^2-y^2$ to lower energies. The arrows mark the involved transitions. See text for discussion.
}
\label{fig:bandasorbdiff}
\end{figure}

Fig.~\ref{fig:bandsU1p6Jp25} (c) shows the reconstructed bands in the $(\pi,0)$ magnetic itinerant regime for $U=1.6$ eV and $J_H=0.25U$ corresponding to a magnetic moment $m=0.91 \mu_B$. The band reconstruction involves band foldings not only at the Fermi surface but also far from it. 
Gaps with different values open at different points in the Brillouin zone in the bands below, at and above the Fermi level. 
Breaking of the $zx-yz$ degeneracy is observed. Along $\Gamma - X$ new bands cross the Fermi level forming V-shaped pockets referred to in the literature as {\em Dirac pockets}. 

When moving towards the orbital differentiated phase with increasing interactions or decreasing doping, the magnetic moment increases and the band reconstruction becomes more complex. Bands become flatter and shift away from the Fermi level, being this effect more accused for the $yz$ and $xy$ orbitals. Spectral weight of these two orbitals shifts partially to higher energies above the Fermi level as these orbitals become half filled.  As shown in Fig.~\ref{fig:bandasorbdiff} (a) for $U=1.6$ eV and $J_H=0.25U$, in the itinerant regime the $yz$ band at $\Gamma$ slightly above the Fermi level (red) lies below a band with dominant $x^2-y^2$ orbital content (black). On the contrary for $U=1.8$ eV, Fig.~\ref{fig:bandasorbdiff} (b), in the orbital differentiation regime the order of these two bands is reversed, driving $yz$ to half-filling. This feature seems to be a fingerprint of the orbital differentiated regime, at least in all the cases analyzed for our tight binding model\cite{nosotrasprb09} and the one by Graser et al~[\onlinecite{graser09}].

\subsection{Optical conductivity}
\label{subsec:results-optcon}

\begin{figure}
\leavevmode
\includegraphics[clip,width=0.48\textwidth]{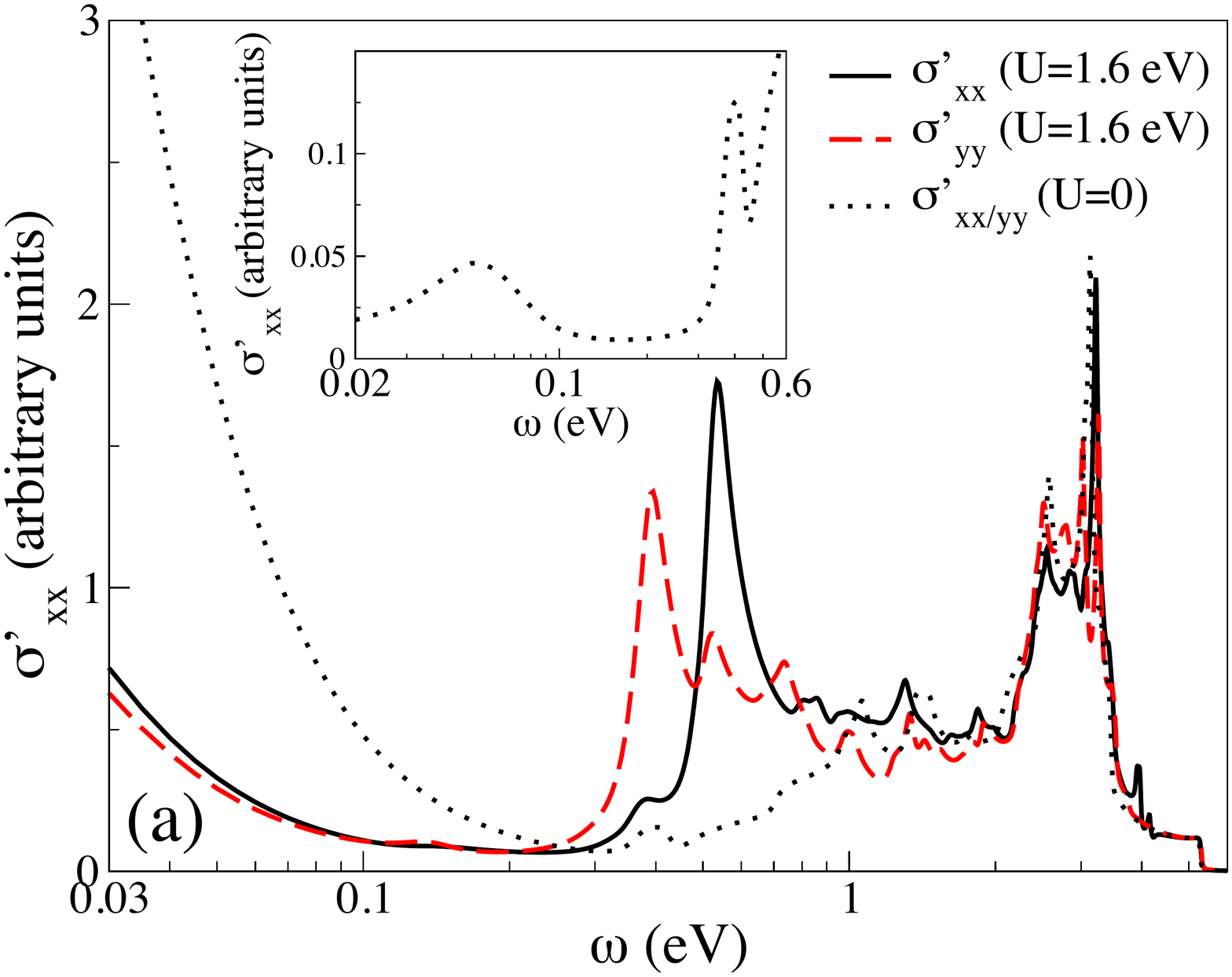}
\includegraphics[clip,width=0.48\textwidth]{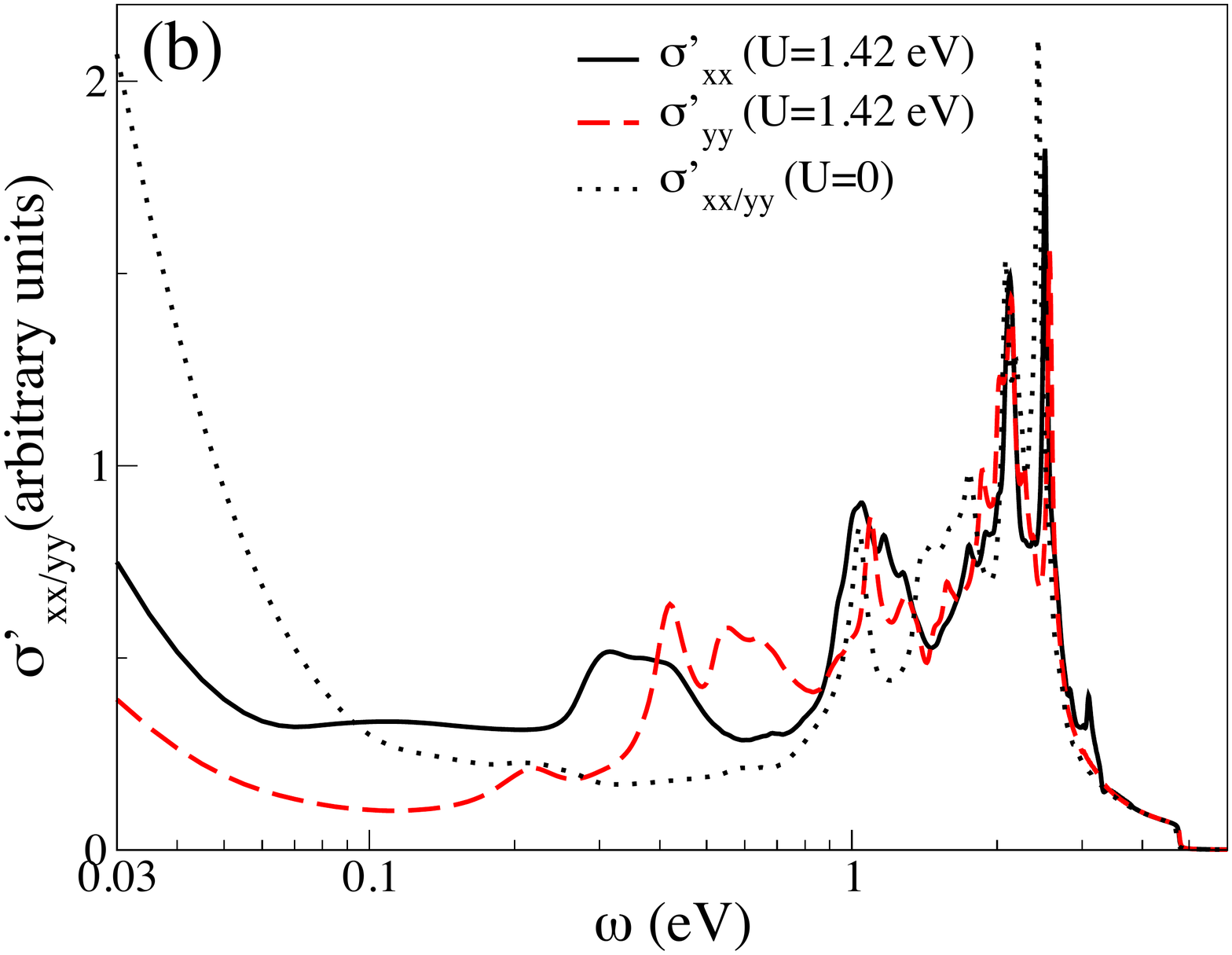}
\vskip -0.2cm
\caption{
(Color online) Optical conductivity for two different tight-binding models. A logarithmic scale for $\omega$ is used to highlight the low energy region. A scattering rate of $20$ meV is used.
(a) Optical conductivity for our tight binding model. \cite{nosotrasprb09} The dotted line corresponds to the non-magnetic state for $U=J_H=0$ eV.  Due to the tetragonal symmetry $\sigma^\prime_{xx}(\omega)=\sigma^\prime_{yy}(\omega)$.  In solid lines, the optical conductivity in the antiferromagnetic $x$ (black) and ferromagnetic $y$ (red) directions corresponding to $U=1.6$ eV and $J_H=0.25U$ in the $(\pi,0)$ antiferrromagnetic state. The tetragonal symmetry is broken and $\sigma^\prime_{xx}(\omega) \neq \sigma^\prime_{yy}(\omega)$. There is a strong suppression of the Drude weight with magnetism: $\sigma'_{xx,yy} (\omega=0, U=0)=12.0876$,  $\sigma'_{xx} (\omega=0, U=1.6 \,{\rm eV})=2.2855$ and $\sigma'_{yy} (\omega=0, U=1.6 \,{\rm eV})=1.9874$.
Inset: Optical conductivity at low energy in the non-magnetic state excluding the Drude peak. A peak corresponding to an interband transition at very low energies ($\omega \sim 50$ meV) shows up. 
(b)  Same as (a) for the tight binding model of Graser et al Ref.~[\onlinecite{graser09}] with  $U=J_H=0$  in the non-magnetic state and $U=1.42$ eV and $J_H=0.25U$ in the magnetic state. See text for discussion. The Drude weights for the varios cases are: $\sigma'_{xx,yy} (\omega=0, U=0)=6.679$,  $\sigma'_{xx} (\omega=0, U=1.42 \,{\rm eV})=2.778$ and $\sigma'_{yy} (\omega=0, U=1.42 \,{\rm eV})=1.2062$.}
\label{fig:optcondU1p6Jp25}
\end{figure}

The optical conductivity is shown in Fig.~\ref{fig:optcondU1p6Jp25} for the non-magnetic state corresponding to $U=J_H=0$ (dotted lines) and for the magnetic state for $U=1.6$ eV and $J_H=0.25 U$ (solid lines) using our model,\cite{nosotrasprb09} Fig.~\ref{fig:optcondU1p6Jp25}(a), and the one by Graser et al,\cite{graser09} Fig.~\ref{fig:optcondU1p6Jp25}(b). A scattering rate of $20$ meV is assumed in the calculations.

Due to the tetragonal symmetry in the non-magnetic state $\sigma^\prime_{xx}(\omega)=\sigma^\prime_{yy}(\omega)$. Beyond the Drude peak, at low frequencies, interband transitions are clearly visible starting from $0.35$ eV and contribute to the conductivity up to energies above $5$ eV. In the following we focus on the low energy transitions. As marked in Fig.~\ref{fig:bandsU1p6Jp25}(b) and following the squared vertices in Fig.~\ref{fig:optramanvertex}, the peaks in $\sigma_{xx}(\omega)$ and $\sigma_{yy}(\omega)$ around $0.4$ eV originate in transitions at $\Gamma$ involving $3z^2-r^2\rightarrow yz$ and $3z^2-r^2\rightarrow zx$ respectively. Similarly, the shoulders  at $0.45$ and $0.7$ eV in $\sigma_{xx}(\omega)$ come from interband transitions involving $zx\rightarrow xy$ along $Y-\Gamma$ and $yz \rightarrow x^2-y^2$ close to $\Gamma$, but not at $\Gamma$. Symmetry related transitions $yz\rightarrow xy$ along $X-\Gamma$ and $zx \rightarrow x^2-y^2$ close to $\Gamma$, but not at $\Gamma$, give an identical contribution to $\sigma_{yy}(\omega)$. While not easily identifiable in the spectrum, interband transitions between the two hole pockets at $\Gamma$ contribute at frequencies within the Drude peak. This interband transition is very narrow, as can be seen in the inset of Fig.~\ref{fig:optcondU1p6Jp25}(a) where the Drude peak has been subtracted from the optical conductivity. This transition is allowed by the finite $e_g$ content of the $yz/zx$ dominated hole bands.\cite{benfatto2011}

In the $(\pi,0)$ magnetic state $\sigma^\prime_{xx}(\omega) \neq \sigma^ \prime_{yy}(\omega)$, as expected from the tetragonal symmetry breaking. The two curves cross many times through all the frequency range, namely, the sign of the optical conductivity anisotropy is frequency dependent.  
In the itinerant regime, for magnetic moments $\sim 0.9 \mu_B$ similar to the experimental ones, the main effect of magnetism is the partial suppression of the Drude peak, which becomes anisotropic, and the appearance of a magnetic peak, see Fig.~\ref{fig:optcondU1p6Jp25} (a). This peak originates in transitions across a magnetic gap between two anticrossed bands. The position of this peak depends on the magnetic moment and is different for $\sigma_{xx}(\omega)$ and $\sigma_{yy}(\omega)$ because, as it can be seen in Fig.~\ref{fig:optramanvertex}, each conductivity samples a different region in ${\bf k}$-space. 

The magnetic peak in $\sigma_{yy}(\omega)$ originates in the $xy \rightarrow yz$ transition at the electron hole pocket at $X$ along the $X-M$ direction and the folded $yz \rightarrow xy$ transition at the $yz$ hole pocket at $\Gamma$ along $\Gamma-Y$, see Fig.~\ref{fig:bandsU1p6Jp25} (d). On the other hand, the peak in $\sigma_{xx}(\omega)$ is due to a $zx \rightarrow xy $ transition close to $Y$ along $Y-M$. A transition with the same energy happens at the hole pocket in $M$. Thus the peak in $\sigma_{yy}(\omega)$ measures the gap at the electron pocket at $X$ and the hole pocket at $\Gamma$ while the one in  $\sigma_{xx}(\omega)$ measures the gap at the electron pocket at $Y$ which is the same as at the hole pocket at $M$. 

\begin{figure*}
\leavevmode
\includegraphics[clip,width=0.47\textwidth]{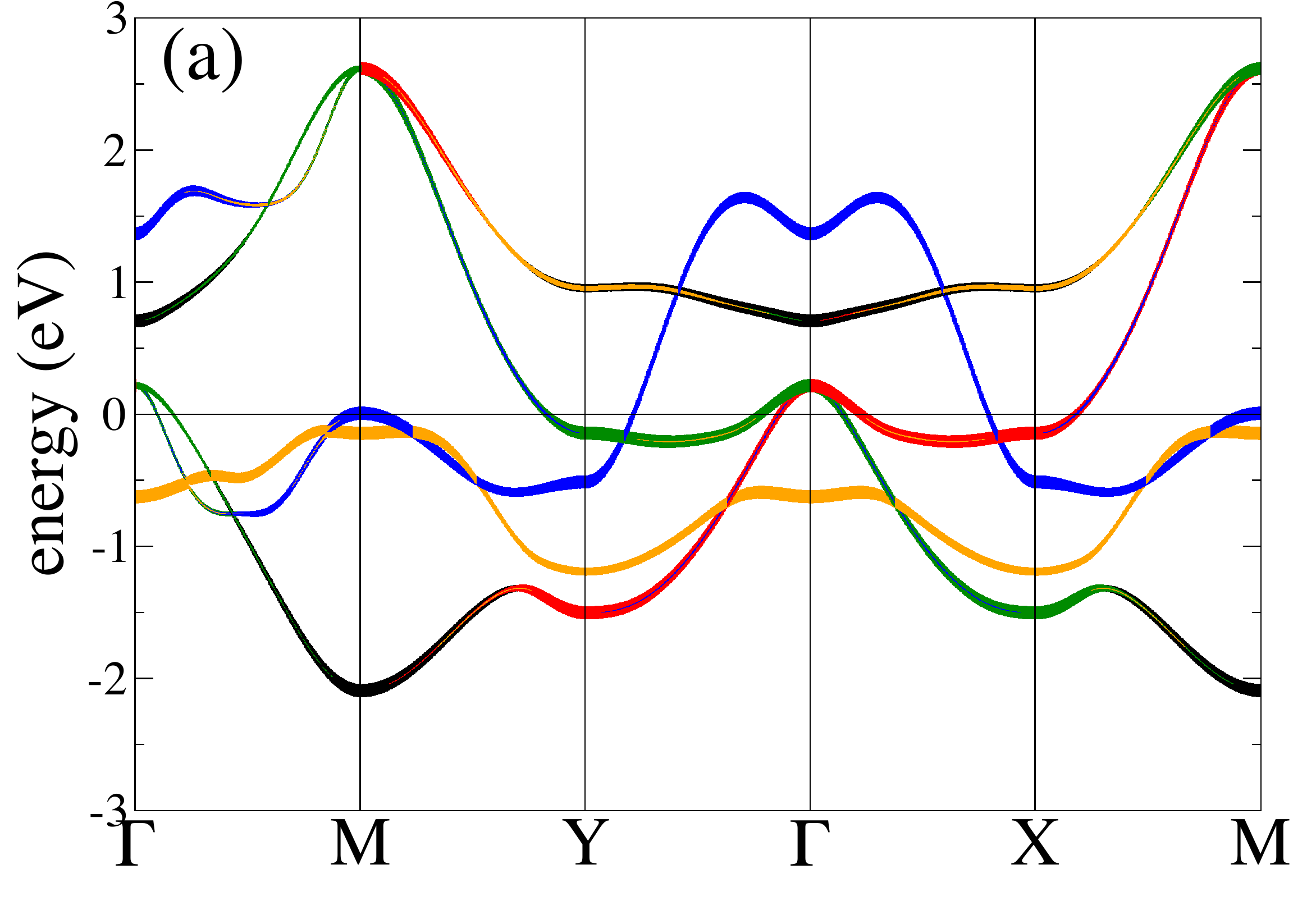}
\includegraphics[clip,width=0.47\textwidth]{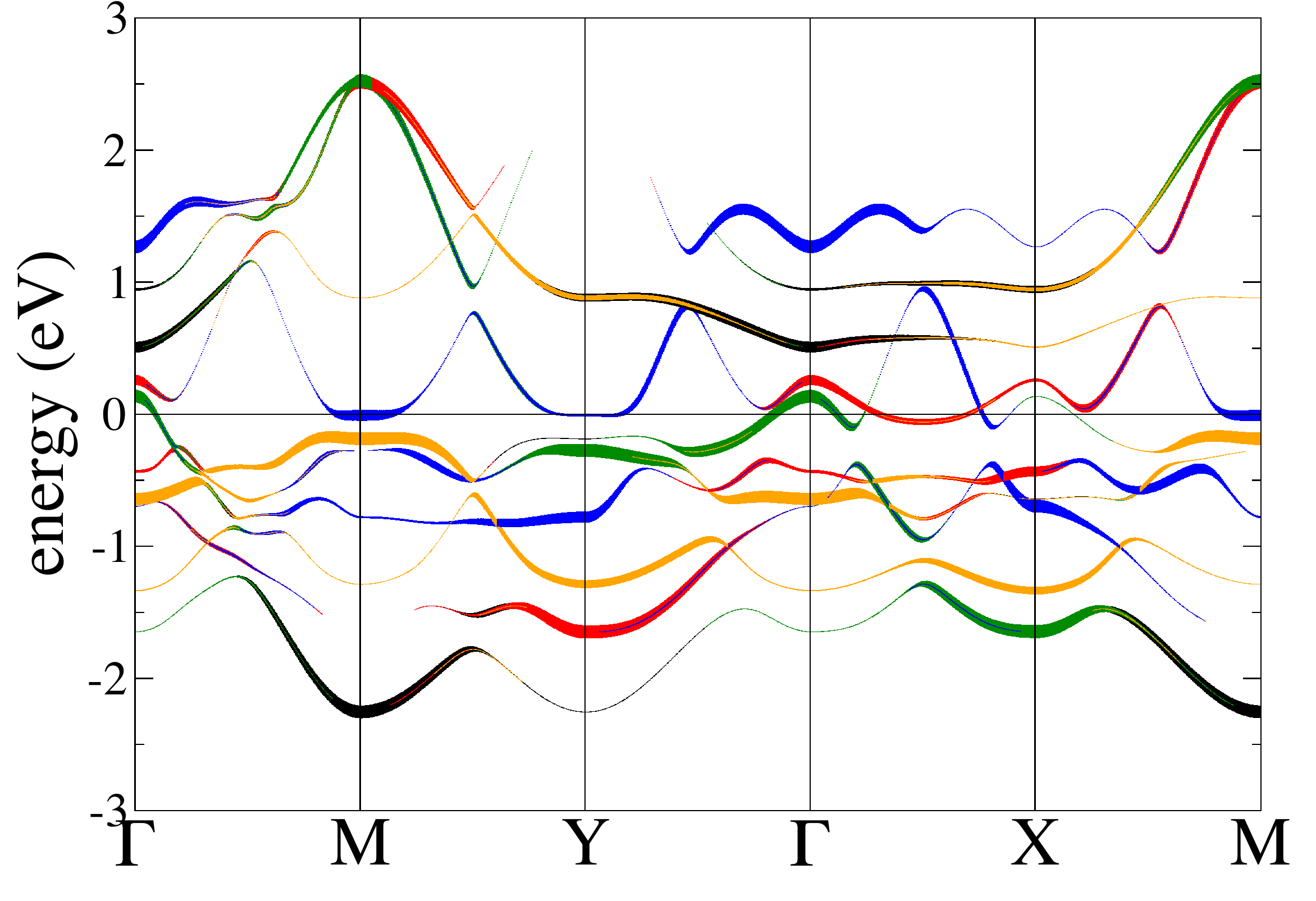}
\includegraphics[clip,width=0.47\textwidth]{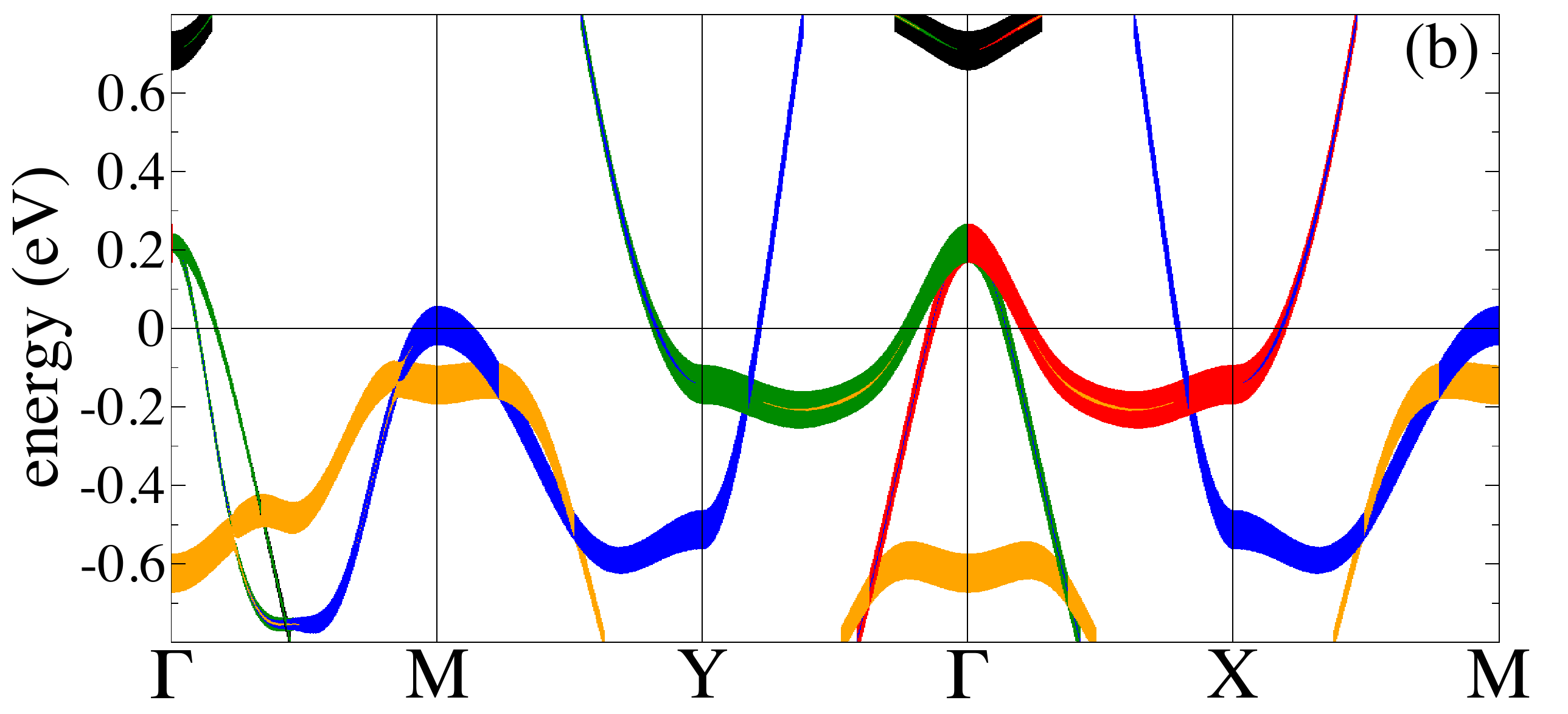}
\includegraphics[clip,width=0.47\textwidth]{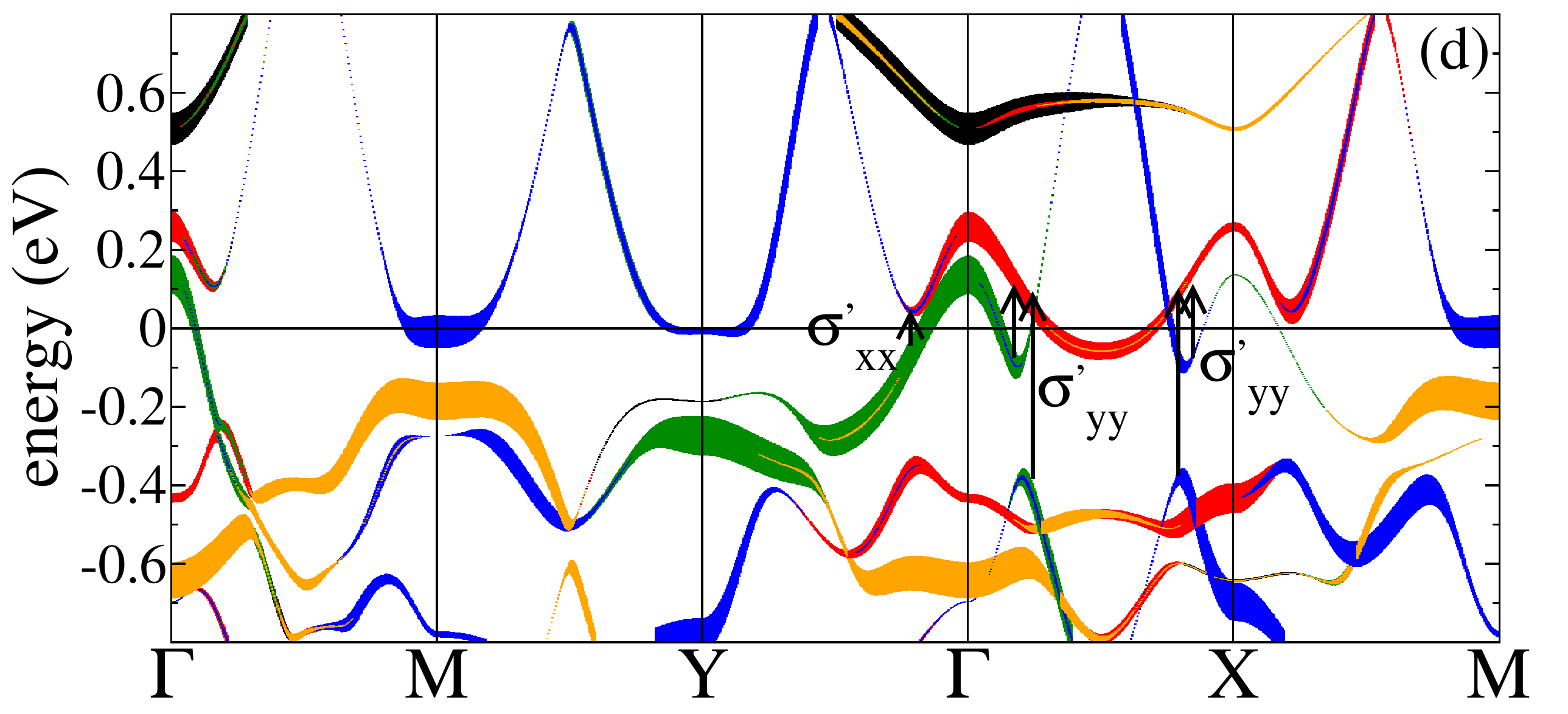}
\vskip -0.2cm
\caption{(Color online) Upper figures: Energy bands for the tight-binding model of Ref.~[\onlinecite{graser09}] in the  non-magnetic (a) and $(\pi,0)$ antiferromagnetic state (c) for $ U=J_H=0$ and $U=1.42$ eV and $J_H=0.25 U$ respectively. Linewidths and colors reflect the orbital content $yz$=red, $zx$=green, $xy$=blue, $3z^2-r^2$=orange and $x^2-y^2$=black. Lower figures: Zoom of the band structures in the upper figures in the energy region close to the Fermi level. (b) The allowed transitions are the same as in Fig.~\ref{fig:bandsU1p6Jp25} (b). \mbox {(d) Transitions} involving a non-folded and a folded band allowed in this case but not in Fig.~\ref{fig:bandsU1p6Jp25} (d), see text. }
\label{fig:bandas-scal} 
\end{figure*}

For the values reported in Fig.~\ref{fig:optcondU1p6Jp25}(a) the magnetic peak in $\sigma_{xx}(\omega)$ appears at larger frequencies than that in $\sigma_{yy}(\omega)$. The gap at the electron pocket at $Y$ depends on the size and height of the hole pocket at $M$ which is very sensitive to the position of the As in the FeAs layer.\cite{vildosola08,nosotrasprb09} Fig.~\ref{fig:optcondU1p6Jp25}(b) shows the optical conductivity corresponding to Graser et al tight-binding model in Ref.~[\onlinecite{graser09}] for $U=1.42$ eV and $J_H=0.25 U$. The magnetic moment is $m=0.91 \mu_B$, as for the values used with our tight-binding model in Fig.~\ref{fig:optcondU1p6Jp25}(a). However the positions of the magnetic peaks are reversed, with the magnetic peak on $\sigma_{xx}$ at a lower frequency than that on $\sigma_{yy}$ in the Graser et al case. The electronic bands corresponding to this model are shown in 
Fig.~\ref{fig:bandas-scal}. The $xy$ hole pocket at $M$ touches the Fermi level in the non-magnetic bands. Even though the bands are similar in the paramagnetic state, compare Fig.~\ref{fig:bandsU1p6Jp25}(a) and Fig.~\ref{fig:bandas-scal}(a), the reconstructed bands in the $Y-M$ direction close to the Fermi level in Fig.~\ref{fig:bandas-scal}(d) are very different from the ones in our model Fig.~\ref{fig:bandsU1p6Jp25}(d). The gaps at the electron pockets differ in both models even though we have chosen parameters such that the magnetic moment is the same. 

In the two cases discussed here the magnetic transitions at both electron pockets are allowed. However, if the minimum of any of the upper folded bands lies below the Fermi level the corresponding transition would be forbidden and the spectrum strongly modified. This happens in Ref.~[\onlinecite{japoneses11}] where a small electron pocket is formed close to $X$ in the $X-M$ directions (and correspondingly close to $\Gamma$ in the $\Gamma-Y$ direction) resulting in the suppression of the $\sigma_{yy}$ magnetic peak observed in our model and highlighted in Fig.~\ref{fig:bandsU1p6Jp25}(d). It can also happen in other models close to $Y$ if the hole pocket at $M$ is below the Fermi level in the non-interacting bands.     
 
 The magnetic peak is the main but not the only signature of magnetism in the optical conductivity at low frequencies.  There are smaller peaks and plateau like structures which originate in transitions close to $\Gamma$ (and equivalently close to $X$) between a folded band and a non-folded band,\cite{japoneses11,gallaisprb11} see Fig.~\ref{fig:transitions}(b). Whether these transitions are allowed depends on the starting tight-binding model and the magnetic moment. For example, in our model, see Fig.~\ref{fig:bandsU1p6Jp25}(d), the gap in the $zx$ hole pocket along $\Gamma-X$ opens below the Fermi level at a value of momentum $k_x$ for which the upper $yz$ band is occupied and the transitions involving these bands at this ${\bf k}$-point are forbidden. On the contrary, in Graser et al model, see Fig.~\ref{fig:bandas-scal}(d), these transitions are allowed and produce the peaks at $0.21$ and $0.55$ eV in $\sigma_{yy}(\omega)$.  
The small contribution to $\sigma_{xx}(\omega)$ below the magnetic peak comes, on the other hand, from $zx \rightarrow xy$ transitions along $\Gamma-Y$. 

\begin{figure}
\leavevmode
\includegraphics[clip,width=0.48\textwidth]{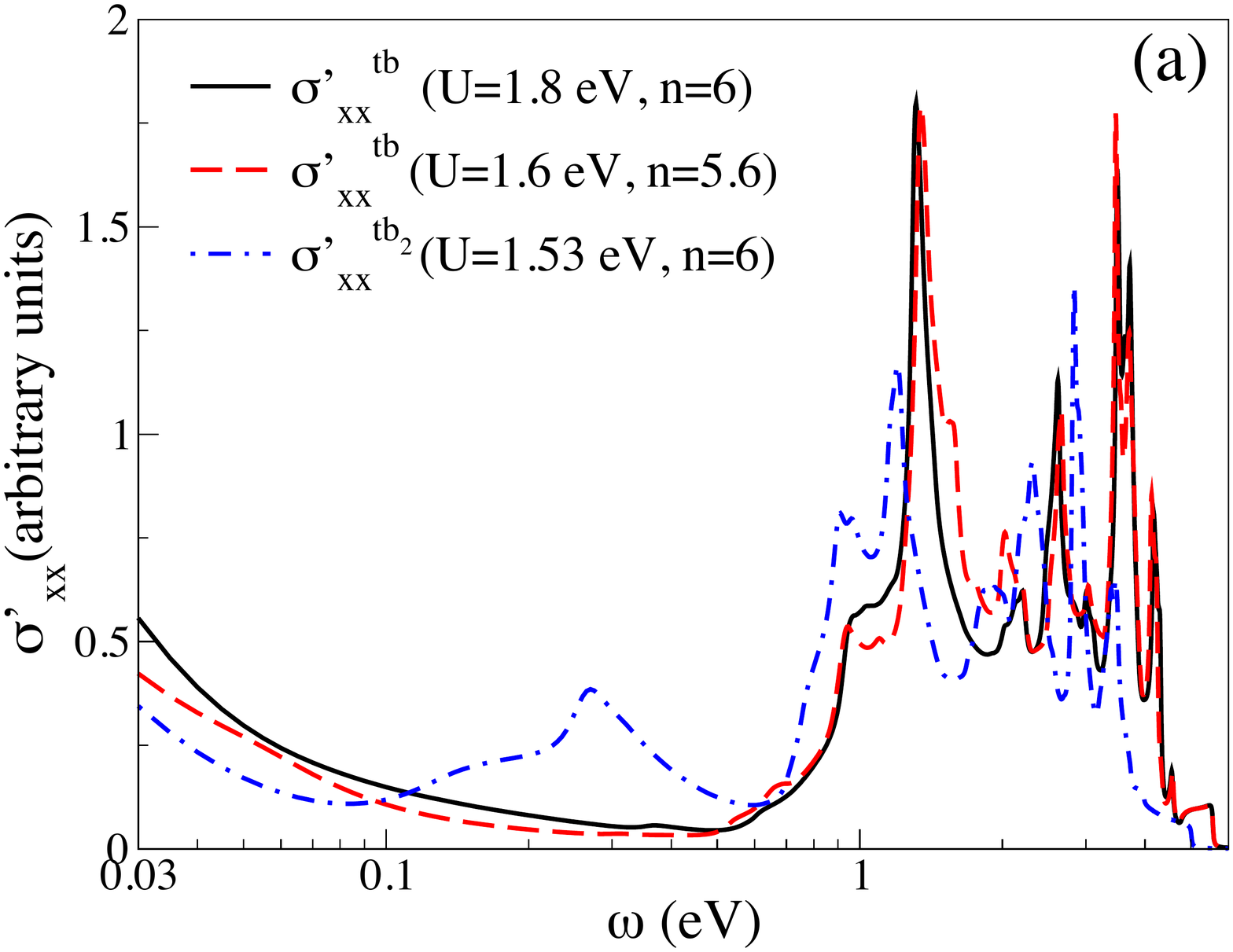}
\includegraphics[clip,width=0.48\textwidth]{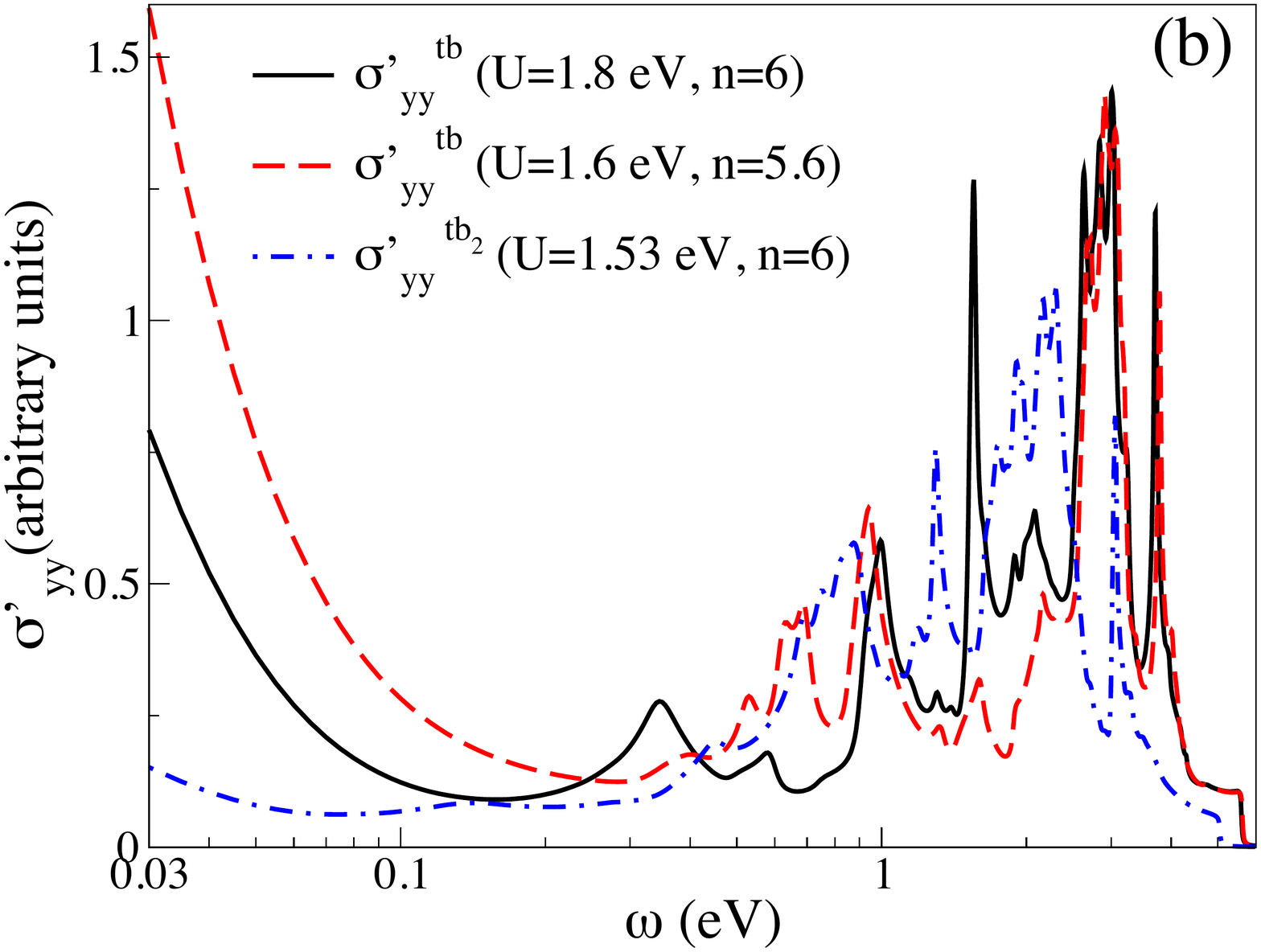}
\vskip -0.2cm
\caption{
(Color online) 
(a) $\sigma_{xx}(\omega)$ (top) and (b) $\sigma_{yy}(\omega)$ in the orbital differentiated region for $U=1.8$ eV and $J_H=0.25U$ (black), $U=1.6$ eV, $J_H=0.25 U$ and electron filling $n=5.6$ (red) using our tight-binding model (labelled tb), and for  
$U=1.53$ eV and $J_H=0.25U$ and $n=6$ (blue) using the Graser et al tight-binding model~\cite{graser09} (labelled tb$_2$). A scattering rate of $20$ meV is used. The Drude weights are $\sigma_{xx}^{\prime tb} (\omega=0, U=1.8 \,{\rm eV}, n=6)=1.6723$,  $\sigma_{xx}^{\prime tb} (\omega=0, U=1.6 \,{\rm eV}, n=5.6)=1.1$ and $\sigma_{xx}^{\prime tb2} (\omega=0, U=1.53 \,{\rm eV}, n=6)=1.05$, $\sigma_{yy}^{\prime tb} (\omega=0, U=1.8 \,{\rm eV}, n=6)=2.5277$,  $\sigma_{yy}^{\prime tb} (\omega=0, U=1.6 \,{\rm eV}, n=5.6)=4.9817$ and $\sigma_{yy}^{\prime tb2} (\omega=0, U=1.53 \,{\rm eV}, n=6)=0.45$}
\label{fig:optcondorbdiff}
\end{figure}

With increasing magnetic moment, in the orbital differentiation region, the shape of the spectrum changes, see Fig.~\ref{fig:optcondorbdiff}, and the spectral weight shifts to higher energies.\cite{japonesesmf09}  The interband transition between the $3z^2-r^2$ band below the Fermi level at $\Gamma$ and the $yz$ band, strongly affected by the orbital differentiation in Fig.~\ref{fig:bandasorbdiff}, is active for $\sigma_{xx}(\omega)$. It produces a step like feature in the spectrum. However because the spectrum is very sensitive to parameters it is not possible to signal an easily identifiable fingerprint of this transition in the spectrum.

\subsection{Raman spectra}

\begin{figure}
\leavevmode
\includegraphics[clip,width=0.48\textwidth]{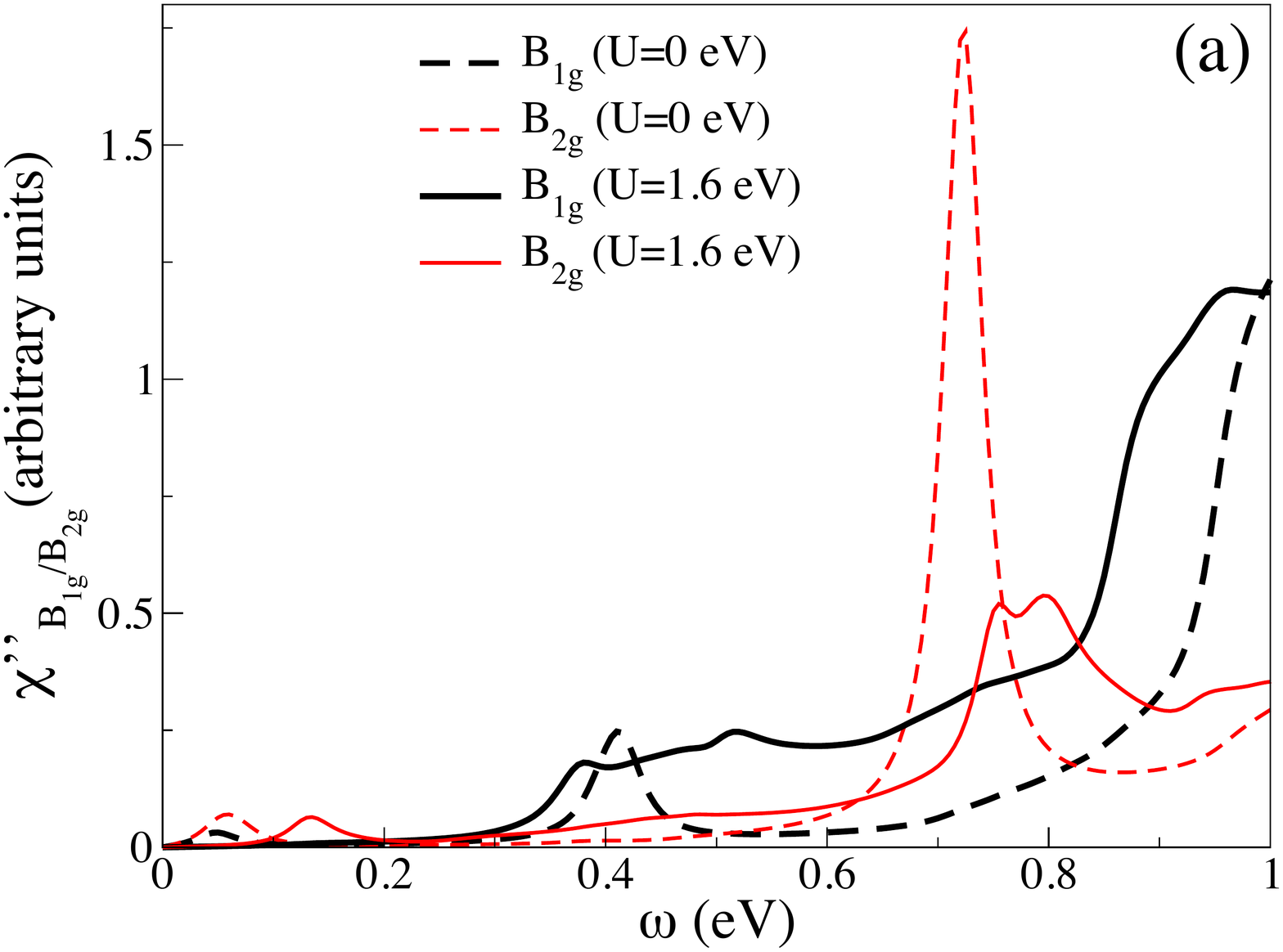}
\includegraphics[clip,width=0.48\textwidth]{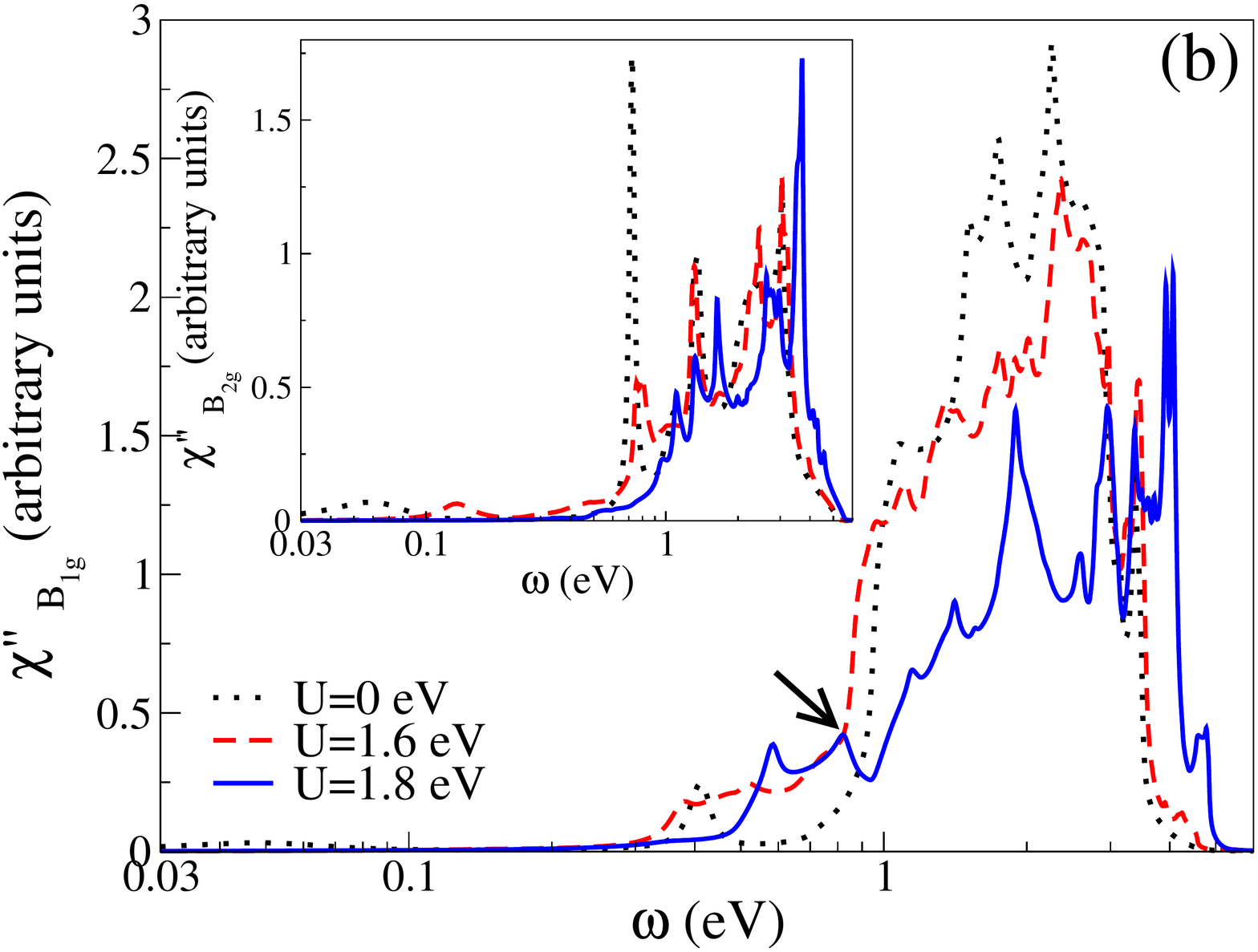}
\vskip -0.2cm
\caption{
(Color online) (a) $B_{1g}$ (black) and $B_{2g}$ (red) Raman spectra. Low frequency Raman spectrum for $U=J_H=0$ in the non-magnetic state (dotted lines) and for $U=1.6$ eV and $J_H=0.25U$ in the itinerant magnetic state (solid lines). (b) Main figure (inset) $B_{1g}$ ($B_{2g}$) Raman spectrum for $U=J_H=0$ (non-magnetic state), $U=1.6$ eV and $J_H=0.25U$ (itinerant magnetic state) and $U=1.8$ eV and $J_H=0.25U$ (magnetic state, orbital differentiated regime) . In $B_{1g}$ the shape of the spectrum changes when entering into the orbital differentiated region. The transition between the $3z^2-r^2$ and $x^2-y^2$ at $\Gamma$ is marked with an arrow. A scattering rate of $20$ meV is used.  
}
\label{fig:raman}
\end{figure}

 The $B_{1g}$ and $B_{2g}$ Raman spectra in the non-magnetic state are plotted in Fig.~\ref{fig:raman}(a). Both symmetries display a peak at very small energies, however the nature of these two peaks is different. In $B_{1g}$ it corresponds to intra-orbital transitions at the electron pockets related to the Drude contribution to the optical conductivity, and it is allowed by the finite scattering rate. In $B_{2g}$ it originates in the interband transition between the $zx$ and $yz$ hole pockets close to $\Gamma$. The rest of the spectrum comes from interband transitions. The strong peak in $B_{2g}$ around $0.7$~eV corresponds to a transition between the $3z^2-r^2$ and the $xy$ hole pockets at $M$. The energy of these hole bands is very sensitive to the As height in the FeAs layer and the transition could be absent if both bands lie below the Fermi level. The peak at $0.4$ eV in $B_{1g}$ originates in an interband transition at $\Gamma$ between the $3z^ 2-r^2$ hole band and the hole pockets with non-negligible $x^ 2-y^ 2$ content. For higher energies there is a large bump in $B_{1g}$. It starts with a step like feature coming from the transition at $\Gamma$ between the $3z^2-r^2$ band below the Fermi level and the $x^2-y^2$ above. 

 In the magnetic state the electron pockets become gapped. The peak at low energies in $B_{1g}$ disappears while peaks at the energy of these gaps appear. $B_{1g}$ samples the two gaps responsible for the magnetic peaks in $\sigma^{\prime}_{xx}(\omega)$ and $\sigma^{\prime}_{yy}(\omega)$. If these two gaps are different enough a two-peak structure should be expected in $B_{1g}$. For the values displayed in Fig.~\ref{fig:raman}(a) these peaks arise at energies comparable to that of the interband transition at $\Gamma$ between the $3z^2-r^2$ band and the hole pocket, also splitted by magnetism, and a wide structure is observed. 
  
  $B_{2g}$ is less sensitive to magnetism, as also seen experimentally.~\cite{gallaisprb11} The low energy peak from the transition between the hole pockets at $\Gamma$ shifts to slightly higher energies. The one at $0.7$ eV acquires a double peak structure and its intensity is suppressed due to the gap opening at the $xy$ hole pocket at $M$. 
Spectral weight appears around $0.3$ to $0.6$ eV due to transitions between a magnetic folded band a non-folded one along $\Gamma-X$ and $\Gamma-Y$ directions. 
 
  $B_{1g}$ samples the $3z^2-r^2 \rightarrow x^2-y^2$ transition at $\Gamma$ whose shape is strongly affected when entering in the orbital differentiated regime with increasing magnetic moment. As shown in  Fig.~\ref{fig:raman}(b) the spectrum changes considerably in this regime. Due to the modification of the $x^2-y^2$ band shape, shown in Fig.~\ref{fig:bandasorbdiff},  the $3z^2-r^2 \rightarrow x^2-y^2$ transition acquires a peak shape instead of a step one. $B_{2g}$ is less affected while the spectral weight is shifted to higher energies. 
 
\section{Drude weight anisotropy}
\label{sec:Drude}
\begin{figure*}
\leavevmode
\includegraphics[clip,width=0.3\textwidth]{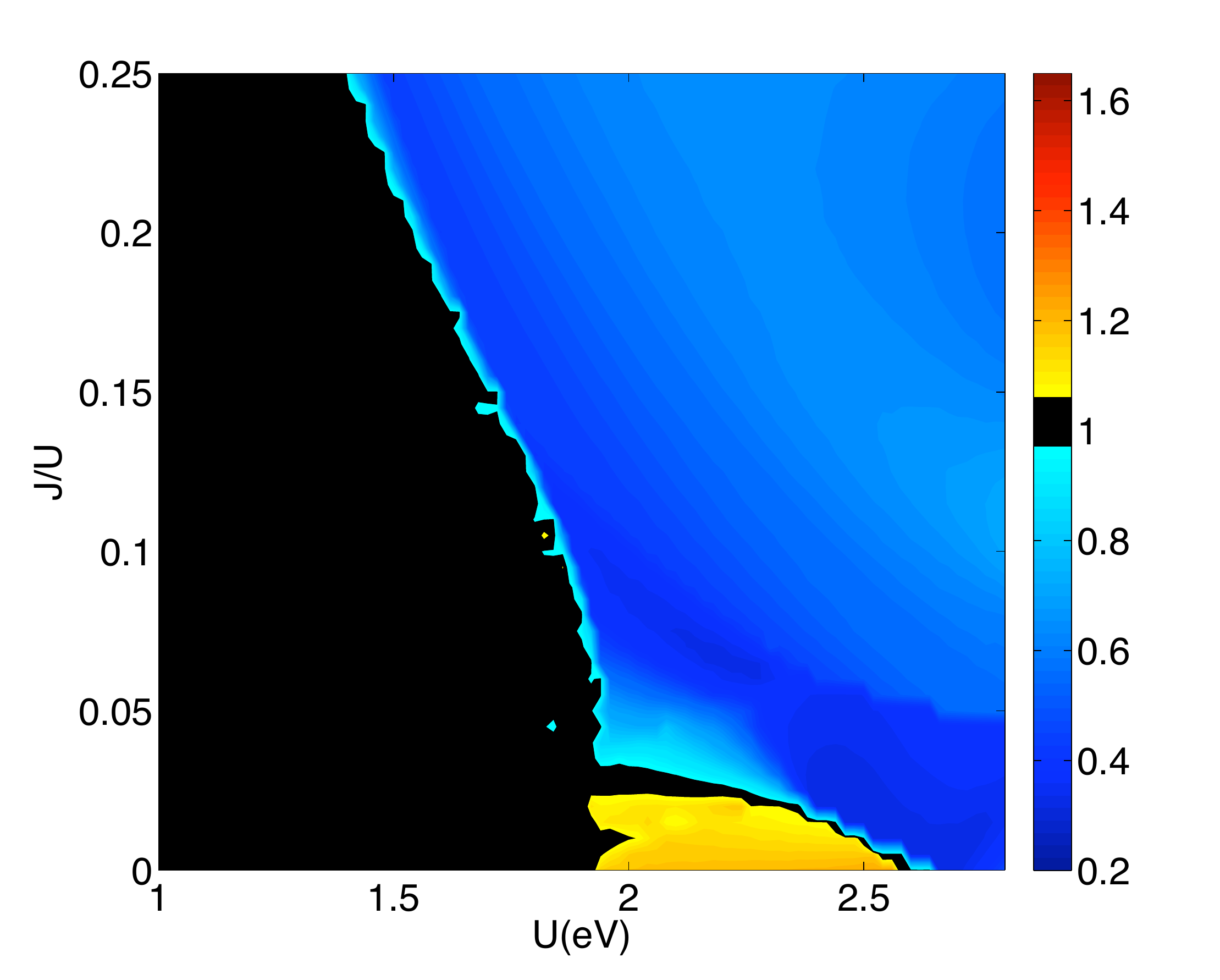}
\includegraphics[clip,width=0.3\textwidth]{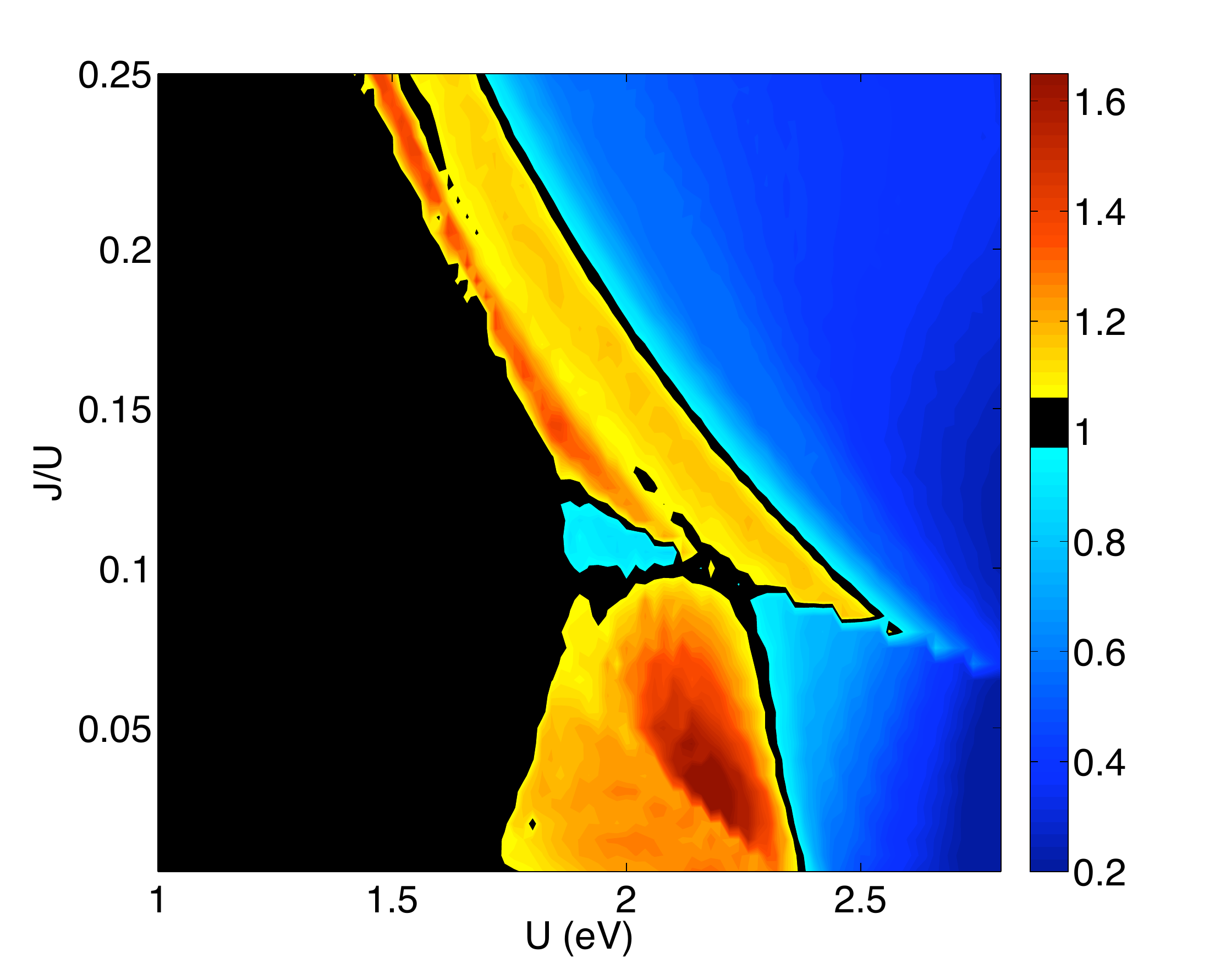}
\includegraphics[clip,width=0.3\textwidth]{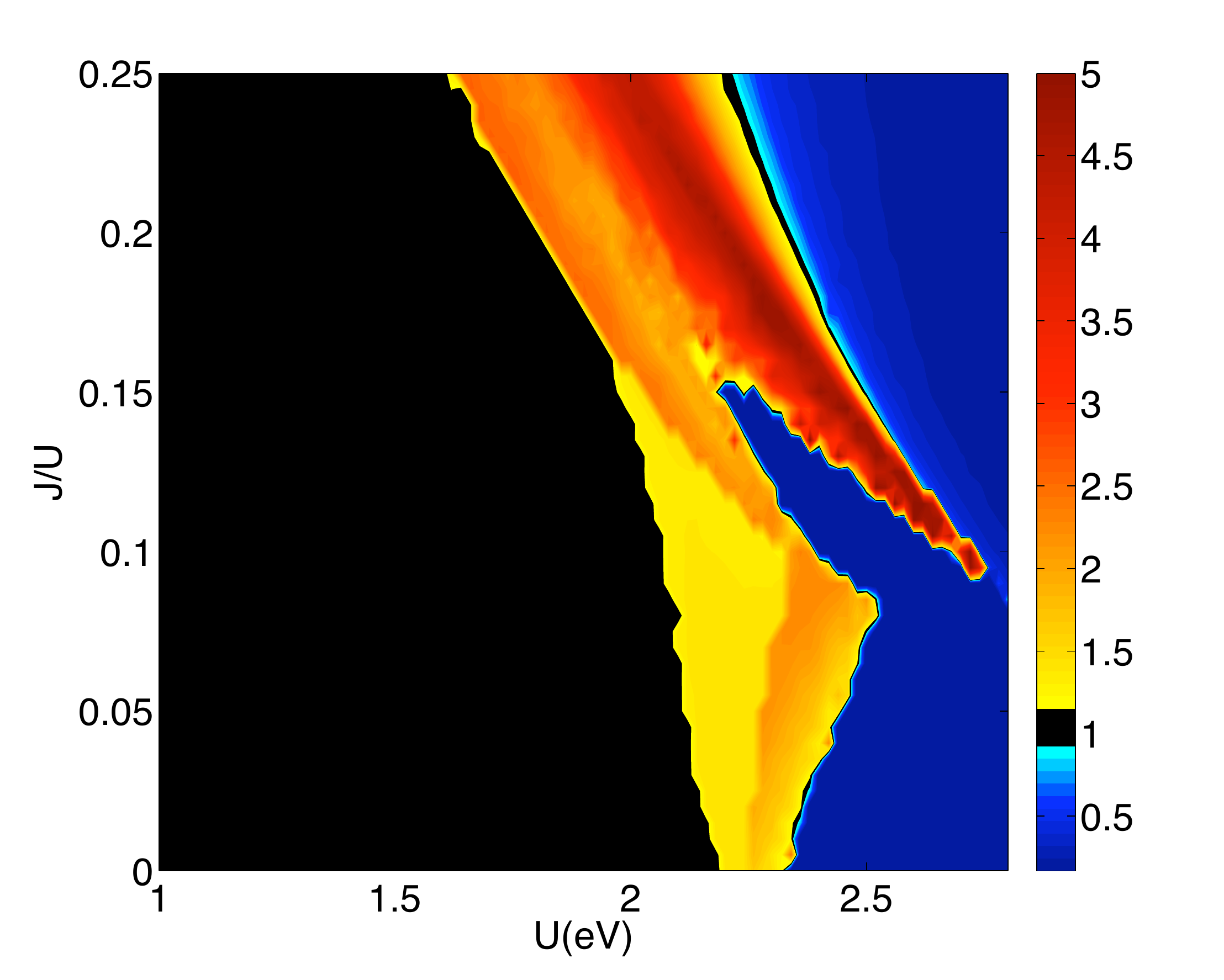}
\vskip -0.2cm
\caption{
(Color online) Drude weight anisotropy $D_x/D_y$ as a function of interactions for three different Fe-As angles, $\alpha_{Fe-As}=33.2^\circ$ as in the slightly squashed tetrahedra found in LaFeAsO (left) $\alpha_{Fe-As}=35.3^\circ$ as in a regular tetrahedra (center) and $\alpha_{Fe-As}= 37.2^\circ$ corresponding to a slightly elongated tetrahedra (right). For small and intermediate values of the interaction $D_x/D_y$ strongly depends on the lattice structure. 
}
\label{fig:ratiodrude}
\end{figure*}
Transport experiments in detwinned samples in the $(\pi,0)$ magnetic state have shown larger resistivity  in the ferromagnetic $y$-direction than in the antiferromagnetic $x$-direction.\cite{chu10-1,mazin10,chen2011,uchida2012,feng2012} The anisotropy is sensitive to disorder and doping. It changes sign in some hole doped samples.\cite{sign-reversal2012} Proposals to explain the anisotropy include orbital ordering,\cite{chen_deveraux10,yin11,lv-phillips2011} integrated or at the Fermi surface,\cite{dagottoprb11} spin nematicity,\cite{Fernandes2011} anisotropic Fermi velocities,\cite{nosotrasprl10-2} and  anisotropic scattering from  disorder in the anisotropic magnetic state.\cite{uchida2012-2}  
Fittings of the Drude peak performed to study whether the anisotropy originates in the Fermi velocities or in the scattering rate are controversial.\cite{degiorgi10,uchida2012-2} The magnetic state is anisotropic. Even if the scattering has a strong effect on the resistivity anisotropy, understanding the anisotropy originated in the reconstructed band structure is necessary. 

Within the mean-field framework above and using the band velocities at the Fermi surface we previously\cite{nosotrasprl10-2} calculated the anisotropy of the Drude weight assuming a regular tetrahedron $\alpha_{Fe-As}=35.3^\circ$. For not very large intraorbital interactions the experimental anisotropy was found. With increasing interactions and magnetic moments the sign of the anisotropy changed. The experimental sign of the anisotropy was concomitant with the smaller values of orbital ordering, discarding orbital ordering as the mechanism for the observed anisotropy.\cite{nosotrasprl10-2}  We proposed that the experimental anisotropy  originated in the topology and morphology of the Fermi surface. Later we argued that in the orbital differentiation regime the system gains kinetic energy in the ferromagnetic $y$-direction inducing a sign change in the anisotropy.\cite{nosotrasprb12-2}

Here we study the sensitivity of the anisotropy to small changes in the lattice.\cite{nosotrasprb09} Fig.~\ref{fig:ratiodrude} shows the Drude ratio $D_x/D_y$ calculated using Eq.~(\ref{eq:drude2}) for three different values of $\alpha$, the angle between the Fe-As bond and the Fe plane: slightly squashed, regular and slightly
elongated tetrahedra with $\alpha_{Fe-As}=33.2^\circ$ (left), $\alpha_{Fe-As}=35.3^\circ$ (center) and $\alpha_{Fe-As}=37.2^\circ$ (right) respectively. On spite of the different expressions used to calculate the Drude weight, the anisotropy map of the regular tetrahedron (middle figure) found here is remarkably similar to that found in our previous work. This similarity supports the interpretation of the Drude weight anisotropy in terms of the Fermi surface velocity.\cite{nosotrasprl10-2}

As expected, for large interactions well into the orbital differentiated
 region\cite{nosotrasprb12-2}  $D_x/D_y<1$, blue color, for the three values of the Fe-As angle.
On the contrary, for small and intermediate interactions the Drude weight
 anisotropy depends strongly on the Fe-As angle. The experimental sign of
the anisotropy $D_x/D_y>1$, in yellow to red colors, is largely
   suppressed in the squashed tetrahedron case. 
This sign of the anisotropy is, on the other hand, preferred in the
 elongated tetrahedron case. The values of the Drude ratio $D_x/D_y$ 
are also much larger in the elongated tetrahedron than in the regular 
one. $D_x/D_y>1$ happens even in a relatively large area of the phase
 diagram in the orbital differentiated region before it switches sign 
 for larger interactions. We believe the values of the interactions
  relevant for iron pnictides are in the region of the phase 
  diagram with anisotropy  sensitive to the Fe-As angle. Such a
   dependence suggests that at least at the level of the reconstructed band structure the resistivity anisotropy is not  a 
   robust fingerprint of the underlying electronic state.

\section{Discussion}
\label{sec:discussion}
In this work we have shown the potential of using the velocity and Raman vertices to disentangle the orbital degree of freedom in the optical conductivity and Raman spectrum of multi-orbital systems. The idea follows the use of Raman vertices in cuprates to differentiate nodal and antinodal regions in $\bf k$-space. The $\bf k$-dependence of the vertices depends on the symmetry of the orbitals involved in the transition. 

We have applied this method to interacting five orbital models for iron superconductors defined in the one-iron unit cell. The optical conductivity and Raman spectra in the magnetic and non-magnetic states have being calculated with the band structure of these five orbital models treated at the mean field level. These mean field bands do not include the renormalization or finite lifetimes due to interactions. Consequently, quantitative agreement between the calculated spectrum and the experiments is not expected. On the other hand, the vertex analysis  is valid independently of the approximation used to calculate the bands and valuable qualitative information can be obtained. 

We have seen that interband transitions involving one or both hole pockets at $\Gamma$ contribute to the optical conductivity in the far and mid-infrared. For non-negligible scattering rates it can be difficult to separate these transitions from the Drude peak. These results confirm previous calculations\cite{benfatto2011} which alerted against using the extended Drude model to analyze the optical spectrum of iron superconductors.\cite{qazilbash09,barisic-dressel2010}
A feature in the optical spectrum of K-doped Ba-122 at frequencies 50-250 cm$^{-1}$ has been interpreted in terms of a pseudogap\cite{kwon2012,lobo2012} precursor of the superconducting gap. Previous works had measured similar features and discussed them in terms of interband transitions\cite{vanHeumenEPL10} or localized states.\cite{lobo2010} Our calculations show that the transition between the two hole pockets at $\Gamma$ is optically active but remains hidden below the Drude peak, see also Ref.~\onlinecite{benfatto2011}. As it is allowed only in a very small region of $\bf k$-space it gives a very narrow contribution to the optical conductivity. Experimentally it could show up at frequencies comparable to that of the superconducting gap, so its presence should be considered when discussing the spectrum. We note that this transition is active in the $B_{2g}$ symmetry (in the one-Fe unit cell). This fact could help clarify the nature of the observed 50-250 cm$^{-1}$ feature in the optical conductivity. \cite{kwon2012,lobo2012}

Experiments in the magnetic state show a suppressed conductivity below $\sim 700$ cm$^{-1}$ and the appearance of a peak around $1000$ cm$^{-1}$~[\onlinecite{hu-wang08,pfuner09,qazilbash09,hu-wang09,dong-wang10,uchida2011,degiorgi10}]. Bump like features show up around $350$ cm$^{-1}$. In detwinned samples the Drude peak is anisotropic, $\sigma^\prime_{xx}(\omega) > \sigma^\prime_{yy}(\omega)$ at low energies but the anisotropy reverses at higher energies. \cite{degiorgi10,uchida2011,uchida2012-2} The Raman spectra show a similar suppression and a bump in all the symmetries, and a peak only in $B_{1g}$ symmetry in the one-Fe unit cell ($B_{2g}$ in the FeAs unit cell).\cite{gallaisprb11}

As discussed in Sec.~\ref{subsec:results-optcon}, in the itinerant regime the magnetic peaks in $\sigma^\prime_{xx}(\omega)$ and $\sigma^\prime_{yy}(\omega)$ respectively sample the gaps opened at the electron pocket at $Y$ and $X$ (folded in the hole pockets at $M$ and $\Gamma$) via a transition between two folded anticrossing bands. The difference between peaks thus measures the electron pockets gaps anisotropy and is not a consequence of orbital order.\cite{uchida2011} The experimental sign of the anisotropy has been reproduced previously in several  theoretical works~\cite{japoneses11,valenti10,dagottoprb11,degiorgi2012,yin11} but had not been explained in these terms. This anisotropy and even the possibility that any of these transitions is forbidden is sensitive to details of the underlying band structure, see Sec.~\ref{subsec:results-optcon}. 
We have shown that the anisotropy of the Drude weight due to the band reconstruction in the magnetic state is also very sensitive to small changes in the lattice structure.

The low energy $B_{1g}$ Raman spectrum samples the electron pockets. These become gapped in the magnetic state, hence the $B_{1g}$ spectrum shows peaks at the energy of the corresponding gaps. These peaks are expected around the same energies as observed in $\sigma^{\prime}_{xx}(\omega)$ and $\sigma^{\prime}_{yy}(\omega)$. If these gaps are close enough in magnitude a single peak instead of two would show up in the experimental $B_{1g}$ spectrum. Some care is required when interpreting the spectrum as an interband transition between $3z^2-r^2$ band and the hole pockets at $\Gamma$, active in $B_{1g}$ could be close in energy. This interband transition is also affected by magnetism, especially because it involves the hole pockets which stop being degenerate. Due to the orbital symmetry $B_{2g}$ does not sample the magnetic gaps at the electron pockets. Therefore we do not expect a peak in $B_{2g}$ at these frequencies.  Our results are compatible with experiments. In the magnetic state $B_{1g}$ shows a peak at an energy similar to the one at which the peaks in $\sigma^{\prime}_{xx}(\omega)$ and $\sigma^{\prime}_{yy}(\omega)$ are observed while there is no peak in $B_{2g}$ at this energy.

The bump like features  around $350$ cm$^{-1}$ in optical conductivity come most probably from transitions between a folded and a non-folded band, see also Ref.~\onlinecite{japoneses11}.  Even if experimentally bumps appear at similar energies in $\sigma^{\prime}_{xx}(\omega)$ and $\sigma^{\prime}_{yy}(\omega)$, following the vertex analysis we believe that they originate in different regions of $\bf k$-space.  Low energy transitions along $\Gamma-X$ contribute to $\sigma^{\prime}_{yy}(\omega)$ and those along $\Gamma-Y$ contribute  to $\sigma^{\prime}_{xx}(\omega)$. Given the anisotropy along these two directions, the bumps at $\sigma^{\prime}_{xx}(\omega)$ and $\sigma^{\prime}_{yy}(\omega)$ are not expected to show equal spectra what it is compatible with experiments.\cite{uchida2011}

$B_{2g}$ samples partially the excitations that we have previously assigned to the bumps in $\sigma^{\prime}_{xx}(\omega)$ and $\sigma^{\prime}_{yy}(\omega)$. Thus the presence of the bump in experiments in this symmetry is consistent within our expectations. On the other hand we do not expect these excitations  to be active in $B_{1g}$. The feature observed experimentally in $B_{1g}$ around these energies should have different origin. 

In the orbital differentiation region shown in the magnetic mean field phase diagram, see Fig. \ref{fig:phasediagram}, the spectrum is strongly modified with a general shift to higher energies with no clearly identifiable feature except for the $B_{1g}$ signal. For $yz$ to become a half-filled gapped state at the orbital differentiation transition there is a shift of the $yz$ orbital to higher energies that modifies the $3z^2-r^2 \rightarrow x^2-y^2$ transition at $\Gamma$ active in $B_{1g}$. As a result the step feature typical of the itinerant regime becomes a peak. 

Finally the  $3z^ 2-r^ 2 \rightarrow xy$  interband transition at $M$ is active in $B_{2g}$ Raman symmetry what could help clarify whether any of these bands cross the Fermi level and complement photoemission measurements.

We acknowledge conversations with Y. Gallais, L. Degiorgi, A. Millis, E. Capellutti, S. Ciuchi, A. Kemper, C. Bernhard and D. Baeriswyl. We acknowledge funding from \mbox {Ministerio} de Econom\'ia y Competitividad through Grants No. FIS 2008-00124, FIS 2009-08744,  FIS 2011-29689, and from CSIC through Grants No. PIE-200960I033 and PIE-200960I180.

\bibliography{pnictides}

\end{document}